\documentclass[10pt]{article}

\usepackage[colorlinks	= true,
			raiselinks	= true,
			linkcolor	= MidnightBlue,
			citecolor	= ForestGreen,
			urlcolor	= ForestGreen,
			pdfauthor	= {Peyman Mohajerin Esfahani},
			pdftitle	= {},
			pdfkeywords	= {},   
			pdfsubject	= {},
			plainpages	= false]{hyperref}
\usepackage[usenames, dvipsnames]{xcolor}
\usepackage{dsfont,amssymb,amsmath,graphicx,fancyhdr,mdframed}
\usepackage{amsfonts,dsfont,mathtools, mathrsfs,amsthm,wrapfig} 
\usepackage{mdframed,ragged2e,subfig}
\usepackage{enumitem}

\allowdisplaybreaks
\usepackage{algorithm}
\usepackage[noend]{algpseudocode}
\usepackage{tcolorbox}
\usepackage[utf8]{inputenc}
\usepackage[T1]{fontenc}

\allowdisplaybreaks
\date{\today}

\newtheorem{thm}{Theorem}[section]
\newtheorem{prop}[thm]{Proposition}
\newtheorem{defn}[thm]{Definition}

\newtheorem{lem}[thm]{Lemma}

\newtheorem{assumption}[thm]{Assumption}

\newtheorem{example}[thm]{Example}
\newtheorem{remark}[thm]{Remark}

\def \l{\left}
\def \r{\right}
\def \a{\alpha}
\def \R{\mathbb{R}}

\def \ol{\underline}

\def \s{\sigma}

\def \der{\operatorname{d}\hspace{-1pt}} 
\def \Der{\operatorname{D}\hspace{-1pt}}

\def \A{\mathcal{A}}

\def \F{\text{face}}

\def \dist{\operatorname{dist}}

\def \ball{\operatorname{ball}}

\def\D{\Delta}
\def\d{\delta}
\def\O{\mathcal{O}}

\def\rank{\mathrm{rank}}

\def\M{\mathcal{M}}

\def \n{\natural}

\def\int{\mathrm{int}}

\def\trace{\mathrm{tr}}
\def\F{\mathrm{F}}

\def\Der{\mathrm{D}}
\def\Int{\mathrm{int}}

\def\effrank{\mathrm{effrank}}

\usepackage{tikz}
\newcommand*\circled[1]{\tikz[baseline=(char.base)]{
             \node[shape=circle,draw,inner sep=.8pt] (char) {#1};}}

\newcommand{\editae}[1]{{\color{black} #1}}

\newcommand{\vertiii}[1]{{\left\vert\kern-0.25ex\left\vert\kern-0.25ex\left\vert #1 
    \right\vert\kern-0.25ex\right\vert\kern-0.25ex\right\vert}} % Three bar norm

\title{Limitations of Implicit Bias in Matrix Sensing:\\  
 Initialization Rank Matters
}

 \author{Armin Eftekhari\thanks{Department of Mathematics and Mathematical Statistics, Umea University, Sweden
   (\texttt{armin.eftekhari@umu.se}).}
 \and Konstantinos Zygalakis\thanks{School of Mathematics, University of Edinburgh, UK
   (\texttt{k.zygalakis@ed.ac.uk}).}
 }

\begin{document}

\maketitle

\begin{abstract}
% We may  {regard} low-rank matrix sensing as a learning problem with infinitely many possible outcomes in which the desired learning outcome is a planted low-rank matrix.
% that achieves zero training error. 
% To find this desired low-rank matrix, for example, nuclear norm   explicitly biases   the learning outcome towards being low-rank.
In matrix sensing, we first numerically
identify the sensitivity to the initialization rank as 
a new limitation of the implicit bias of gradient flow. We will  partially quantify this phenomenon mathematically, where we establish  that the gradient flow of the  empirical risk is  implicitly biased towards low-rank outcomes and successfully learns the planted low-rank matrix, provided that the initialization is low-rank and within a specific ``capture neighborhood''. This capture neighborhood is far larger than the corresponding neighborhood in local refinement results; the former contains all models with zero training error whereas the latter is a small neighborhood of a model with zero test error.  
These new insights enable us to design an alternative algorithm for matrix sensing that complements the high-rank and near-zero initialization scheme which is predominant in the existing literature.

% In the absence of any explicit regularization, we establish here that the gradient flow of the  empirical risk is  implicitly biased towards low-rank outcomes and successfully learns the planted low-rank matrix, provided that the initialization is low-rank and {within a specific ``capture neighborhood''}. This result and the supporting  numerical evidence identifies a new facet of  implicit bias, suggesting that the initialization rank matters.

\end{abstract}

\section{Introduction}\label{sec:intro}
\vspace{-5pt}
% \textbf{To-do: Rewrite abstract, More simulations for the initialization scheme, Find expressions for initialization neighborhood size for the Gaussian case, Add the manifold assumption to the supplementary, Create the supplementary from this document, not the old one. Notation overline to underline for U. New related work.}

In recent years, beyond its traditional role~\cite{davenport2016overview}, the framework of matrix factorization has also served as a means to gain theoretical insight   {into unexplained} phenomena in  neural networks~\cite{gunasekar2017implicit,arora2019implicit,eftekhari2020training}. 
 {As an example}, a trained deep neural network is an overparametrized learning machine with (nearly) zero training error that nevertheless achieves a small   {test} error,
% That is,  a deep neural network  (nearly) interpolates its   {training} data and yet surprisingly avoids overfitting, 
suggesting that the implicit bias of the training algorithm 
% (which is often a variant of the stochastic gradient descent), 
plays a key role in the empirical success of neural networks~\cite{zhang2016understanding,martin2018implicit}.  {A similar pattern has emerged in various other learning tasks, see for example~\cite{soudry2018implicit,belkin2018understand,xie2020weighted,liang2020just}}.

To  {better} understand  {this implicit bias},  {let us specifically consider} low-rank matrix sensing~\cite{davenport2016overview}.
% When the sample size~$m$ is small, 
We may regard matrix sensing as an overparametrized learning problem  {in which} there are potentially infinitely  {many} models with zero training error 
% ways to minimize the training error to zero 
 {that  perfectly interpolate} the  {training} data~$b\in \R^m$.  {That is},  {if $\A:\R^{d\times d}\rightarrow\R^m$ denotes the linear operator in matrix  sensing},  there are potentially infinitely many matrices that are mapped to~$b$ by the linear operator~$\A$. 
 {However}, the desired learning outcome here is   {a planted} \emph{low-rank} matrix~$X^\n\in \R^{d\times d}$   {that satisfies}~$\A(X^\n)=b$. 
%  {Without any explicit regularization to encourage a low-rank outcome, is it possible to recover $X^\n$?}
% We say that a  numerical algorithm is implicitly biased towards $X^\n$ if it recovers $X^\n$ and \emph{only} its initialization might depend 

Does there exist a numerical algorithm that is implicitly biased towards the planted model $X^\n$? By our convention, such an algorithm would successfully recover~$X^\n$, even though \emph{only} its initialization might exploit our prior knowledge that $X^\n$ is low-rank. 
% (beyond the information available in the training data $b$). } 

Under a restricted injectivity assumption on the operator~$\A$,~a first affirmative answer to the above question appeared in \cite{li2018algorithmic}. Informally speaking, this work established that  the trajectory of the 
% gradient 
flow 
\begin{equation}
\dot{U}(t)  = - \nabla \|\A(U(t)U(t)^\top)-b\|_2^2,
\label{eq:flowIntro}
\end{equation}
when initialized at 
\begin{equation}
 U(0) = u_0 \cdot I_d \in \R^{d\times d}, \qquad u_0\approx 0,
 \label{eq:nearlyZero}
\end{equation}
is such that $U(t)U(t)^\top$ spends a long time near the planted low-rank matrix $X^\n$. 
% provided that $u_0$ is very small. 
(Above, $\dot{U}(t) = \der U(t)/\der t$ and $I_d$ is the identity matrix.) 
% This finding is  remarkable because no external force pushes the flow~\eqref{eq:flowIntro} towards $X^\n$.
% low-rank matrices. 
% Loosely speaking,~\cite{li2018algorithmic} established that the flow~\eqref{eq:flowIntro} is implicitly biased towards the planted model~$X^\n$, when initialized \emph{very close} to the origin. 
In this result, initialization \emph{near} the origin is vital, without which there is in general no implicit bias  whatsoever towards $X^\n$! Indeed, we will shortly use a numerical example to illustrate that the flow \eqref{eq:flowIntro} is in general \emph{not} implicitly biased towards $X^\n$, when initialized 
\emph{far} from the origin.  
% This observation allows us to identify the initialization norm as a key factor in the implicit bias of the flow~\eqref{eq:flowIntro}. 
Phrased differently, sensitivity to the initialization norm is a key limitation of implicit bias in matrix sensing. 
% regularization. 
This observation   echoes~\cite{gunasekar2017implicit,arora2019implicit}. 

\begin{figure}%[h]
        \centering
        \includegraphics[width=6.9cm,height=4.5cm]{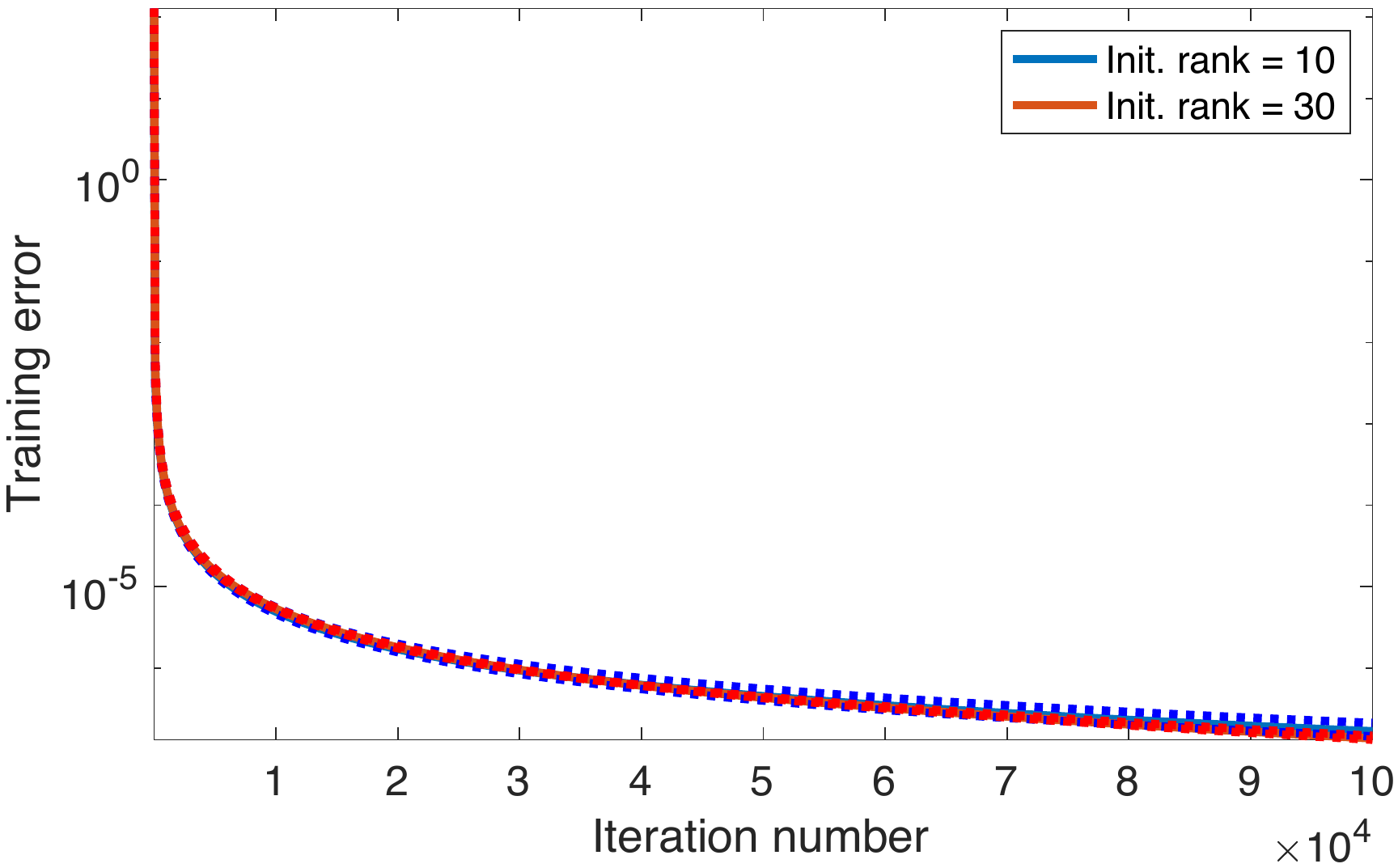}   
        \hfill
        \includegraphics[width=6.9cm,height=4.5cm]{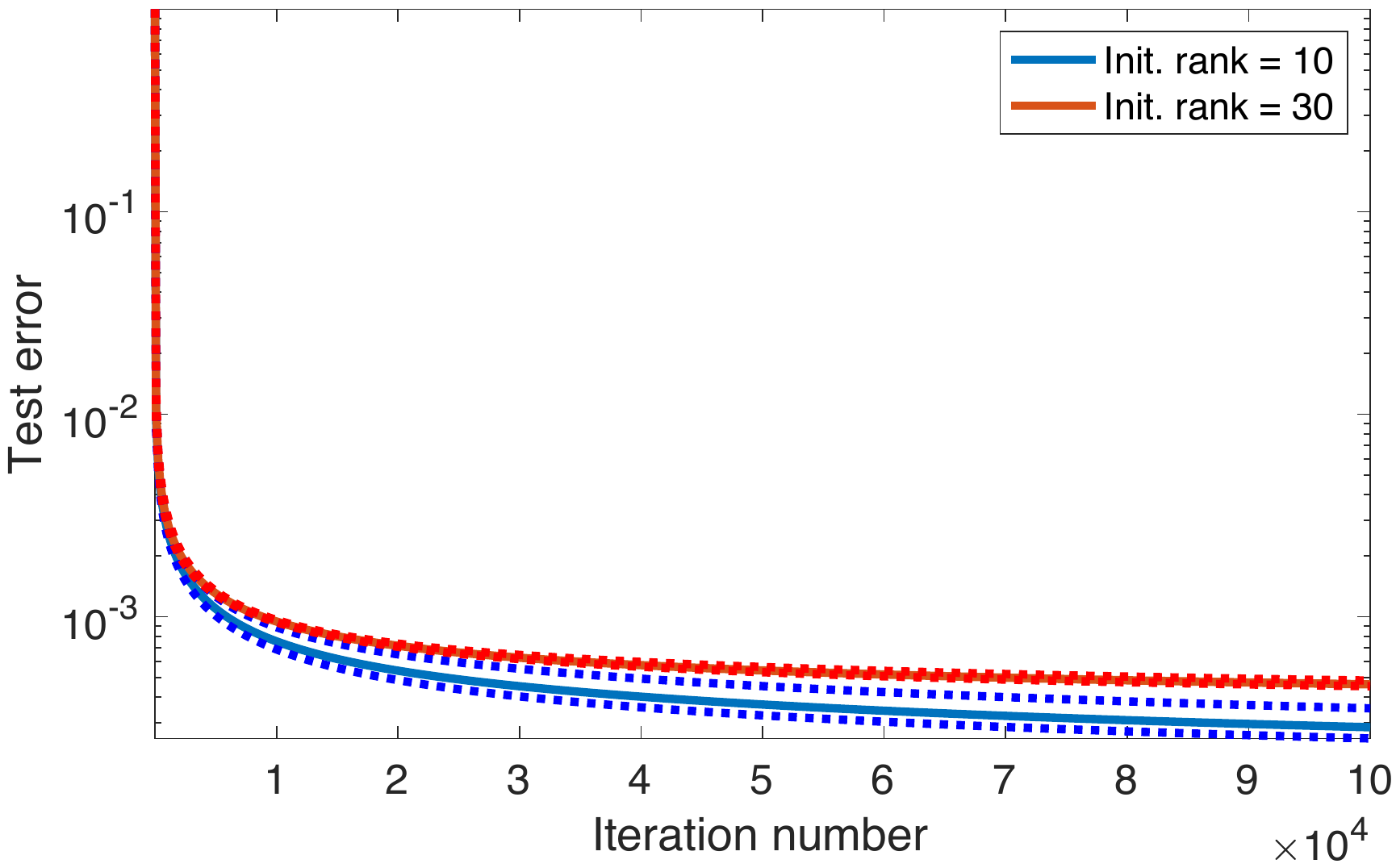}
        \caption{ {With the setup explained in Example~\ref{ex:numEx}, this figure shows the training error (left) and test error (right)  of the gradient descent~\eqref{eq:gradDec}, when initialized near the origin $(\|U_0\|_\F=10^{-3})$. The solid and dotted lines show the average over three trials  $\pm$ half of the standard deviation, respectively.
        Note that the implicit bias of the gradient descent~\eqref{eq:gradDec} towards the planted model~$X^\n$ is (slightly) stronger (and the test error is marginally smaller) when the initialization is low-rank.  
        % The gradient descent displays implicit regularization towards the  planted low-rank matrix only when the initialization norm is small, i.e., when $U_0$ has small singular values.  
        }}
        \label{fig:lowNorm}
    \end{figure}

\begin{figure}%[h!]
        \centering
        \includegraphics[width=6.9cm,height=4.5cm]{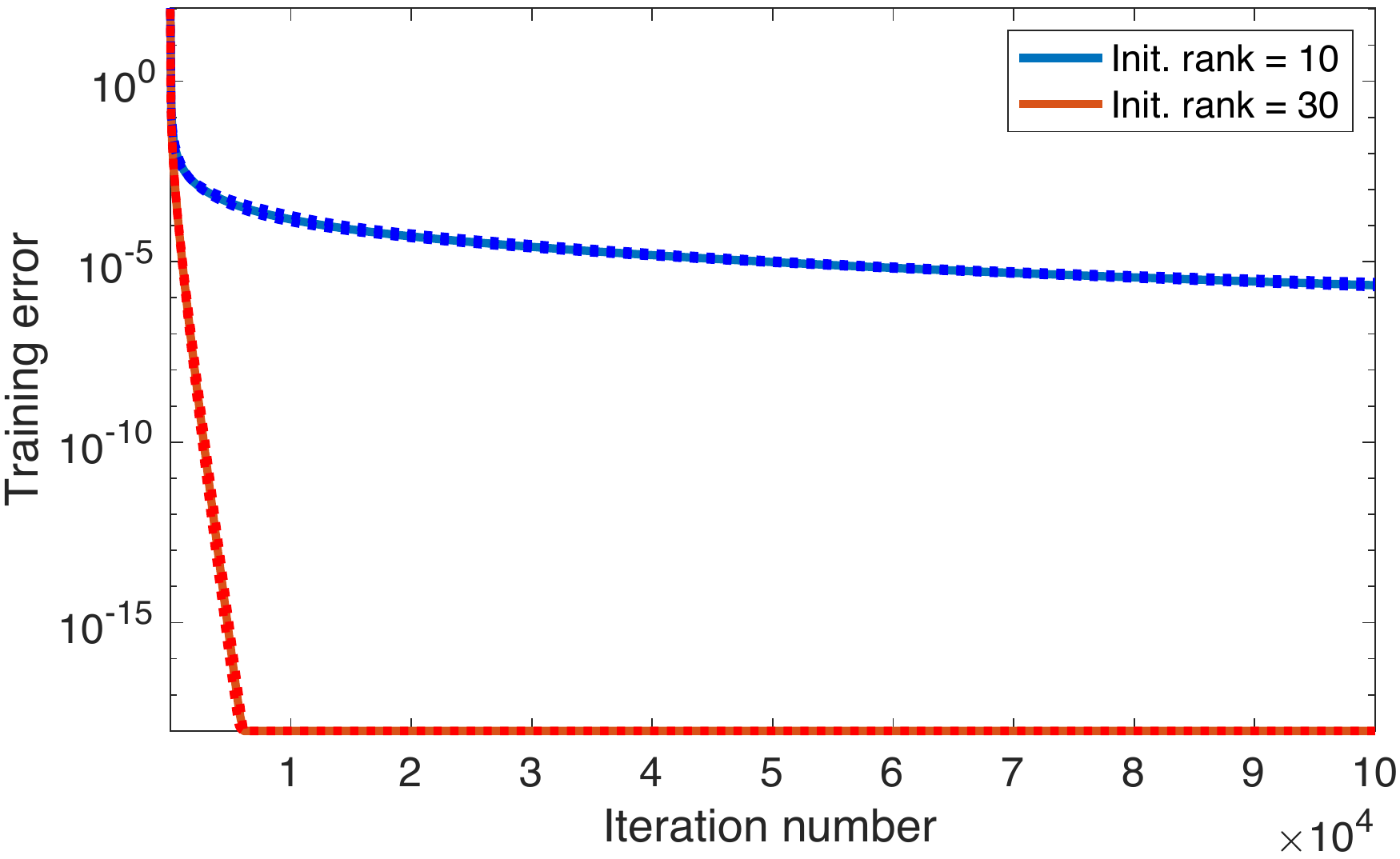}   
        \hfill
        \includegraphics[width=6.9cm,height=4.5cm]{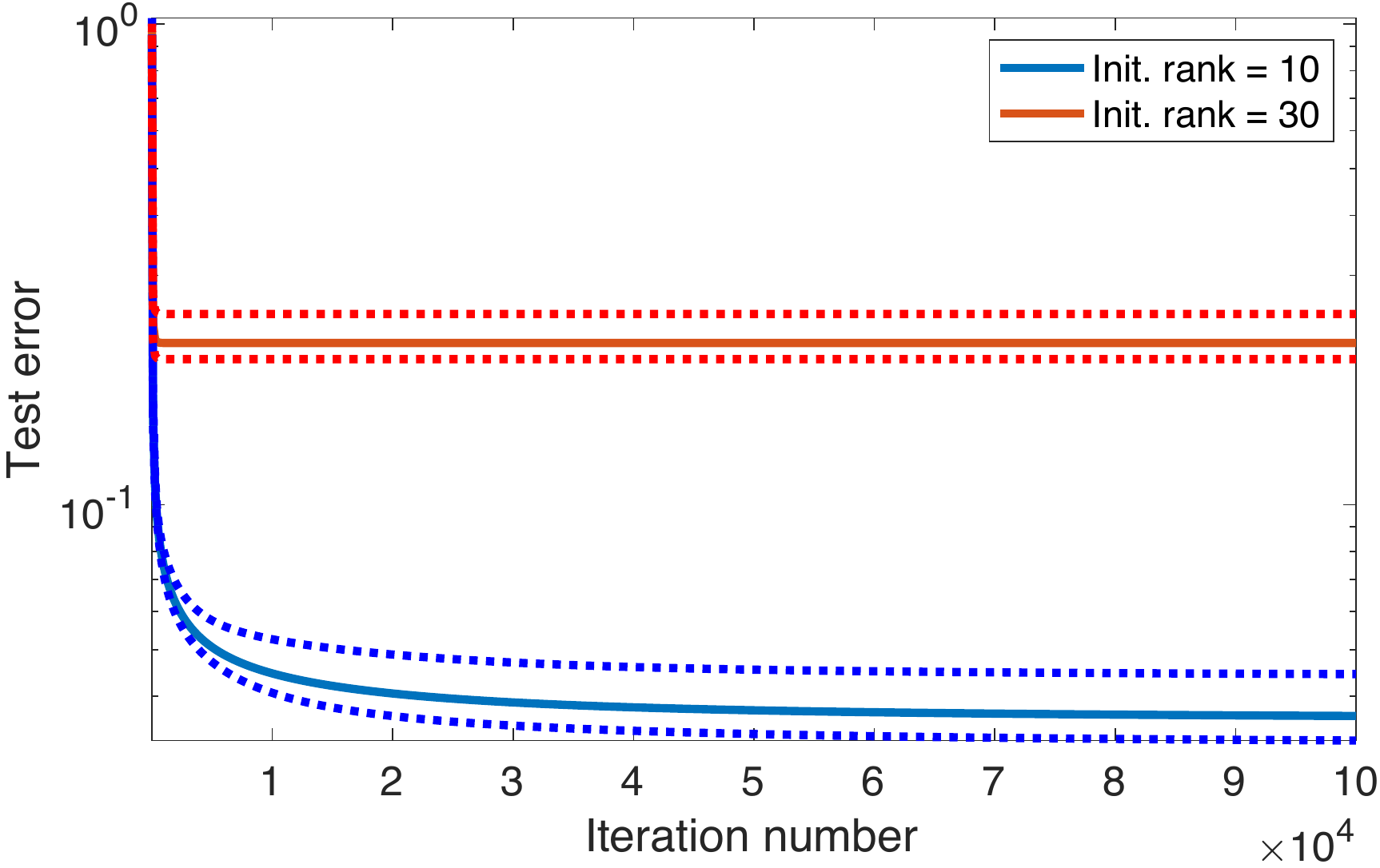}
        \caption{ {
        This figure complements Figure~\ref{fig:lowNorm}.
        With the setup explained in Example~\ref{ex:numEx}, this figure shows the training error (left) and test error (right)  of the gradient descent~\eqref{eq:gradDec}, when initialized further from the origin $(\|U_0\|_\F=1)$. The solid and dotted lines again show the average over three trials  $\pm$ half of the standard deviation, respectively. Note that the implicit bias of the gradient descent \eqref{eq:gradDec} towards the planted model~$X^\n$ is considerably stronger  when the initialization is low-rank. Above, the gradient descent with high-rank initialization terminates after the training error in one of its trials reaches the machine precision.   
        % Gradient descent consistently displays implicit regularization towards the  planted low-rank matrix, regardless of initialization norm, suggesting that it is the initialization rank rather, than its norm, which is chiefly responsible for the implicit regularization of the gradient descent~\eqref{eq:gradDec}. 
        }}
        \label{fig:hiNorm}
    \end{figure}

Our work  pinpoints the  initialization \emph{rank} as another key factor that contributes to the implicit {bias of the gradient flow~\eqref{eq:flowIntro}}. Viewed differently, we identify the sensitivity to the initialization rank as another key limitation of implicit bias in matrix sensing. We will also partially quantify this phenomenon. These insights  later enable us to design an alternative algorithm for matrix sensing that complements the high-rank and near-zero initialization scheme in~\cite{gunasekar2017implicit,arora2019implicit,li2018algorithmic}, see~\eqref{eq:nearlyZero}.
First, let us motivate the role of initialization rank with a small numerical example. 
\begin{example}[\textsc{Initialization rank matters}]\label{ex:numEx}
For $d=30$, we  randomly generate $X^\n\in \R^{d\times d}$ with $\rank(X^\n)=2$ and $\trace(X^\n)=1$. 
For every $i\le m=4\cdot \rank(X^\n)d$, we then populate the upper triangular entries of a symmetric matrix $A_i\in \R^{d\times d}$ with independent Gaussian random variables that have zero mean and unit variance. In this way, we obtain a sensing operator $\A:\R^{d\times d}\rightarrow\R^m$ that maps $X$ to $[\langle A_1,X\rangle, \cdots, \langle A_m, X\rangle]^\top$. Both $X^\n$ and $\A$ are then fixed throughout the rest of this example. As a discretization of the gradient flow~\eqref{eq:flowIntro},
we  implement the gradient descent algorithm
\begin{equation}
U_{k+1} = U_k - \eta \nabla \|\A(U_kU_k^\top) - b\|_2^2,
\label{eq:gradDec}
\end{equation}
with the learning rate of~$\eta = 10^{-4}$ and various choices for the initialization~$U_0\in \R^{d\times d}$. 

We specifically compare the implicit bias of the flow \eqref{eq:flowIntro} towards $X^\n$, when initialized near, far and very far away from the origin: 
Figure~\ref{fig:lowNorm} shows the training error $\|\A(U_kU_k^\top) - b\|_2^2$ and the test error $\|U_kU_k^\top - X^\n\|_\F^2$ for a generic initialization $U_0\in \R^{d\times d}$ that is near the origin $(\|U_0\|_\F = 10^{-3})$, and for two values of $\rank(U_0)$, each averaged over three trials.  Figure~\ref{fig:hiNorm} shows the training and test errors for a generic  initialization~$U_0$ that is further away from the origin $(\|U_0\|_\F = 1)$, and for two values of~$\rank(U_0)$, each averaged  over three trials. Lastly, Figure~\ref{fig:break} corresponds to initialization very far from the origin $(\|U_0\|_\F=10^3)$. 

% As suggested by~\cite{gunasekar2017implicit,arora2019implicit,li2018algorithmic}, 
We observe in Figures~\ref{fig:lowNorm}-\ref{fig:break} that the implicit bias of the gradient descent
% ~\eqref{eq:gradDec} 
{towards~$X^\n$} gradually disappears (and the test error gradually increases) as the initialization norm~$\|U_0\|_\F$ increases. This observation identifies initialization norm as a key factor that contributes to the implicit bias of gradient descent in matrix sensing, echoing the findings of~\cite{gunasekar2017implicit,arora2019implicit,li2018algorithmic}.

% In particular, when the initialization is high-rank, gradient descent is in general \underline{not} implicitly biased towards $X^\n$, see Figure~\ref{fig:hiNorm}. 

% However, regardless of the initialization norm, we also  observe in

An equally remarkable pattern that also emerges in Figures~\ref{fig:lowNorm}-\ref{fig:break} is the sensitivity of implicit bias to the initialization rank. In each figure, we observe that the gradient flow~\eqref{eq:gradDec}, when initialized low-rank, consistently outperforms its high-rank counterpart and displays a stronger implicit bias towards the planted model $X^\n$. That is, a low-rank initialization consistently achieves a smaller test error compared to its high-rank counterpart. The difference is more pronounced in Figures~\ref{fig:hiNorm} and~\ref{fig:break}. From another perspective, the initialization rank is another limitation of implicit bias in matrix sensing. Particularly,  when the initialization is high-rank, in general the gradient descent~\eqref{eq:gradDec} is  \underline{not} implicitly biased towards the planted model $X^\n$, see Figure~\ref{fig:hiNorm}. These patterns are typical across parameters. 
% and persistent. 
% These observations persist across various choices of the parameters.
% and the MATLAB code is available upon publication. 

% Figures~\ref{fig:lowNorm} and~\ref{fig:hiNorm} that the gradient descent~\eqref{eq:gradDec}, with a low-rank initialization, always displays a stronger implicit bias towards the planted model~$X^\n$ (and achieves a smaller test error), when compared to a high-rank initialization.
% \hfill$\blacksquare$
% The red curve in Figure~\ref{fig:hiRank} thus echoes the findings of~\cite[Section 6]{li2018algorithmic}, i.e., the test error is small (and gradient descent thus displays implicit regularization towards $X^\n$), provided that~$\|U_0\|_\F$ is small. 
% On the other hand, Figure~\ref{fig:lowRank} shows the training and test errors for a generic rank-$3$ initialization~$U_0$ with variable~$\|U_0\|_\F$, averaged over three trials. Consistently, a low-rank initialization~$U_0$ leads to a small test error \emph{regardless} of $\|U_0\|_\F$.
\end{example}

\subsection{Contributions}\label{sec:contributions}
\begin{figure}%[h!]
        \centering
        \includegraphics[width=6.9cm,height=4.5cm]{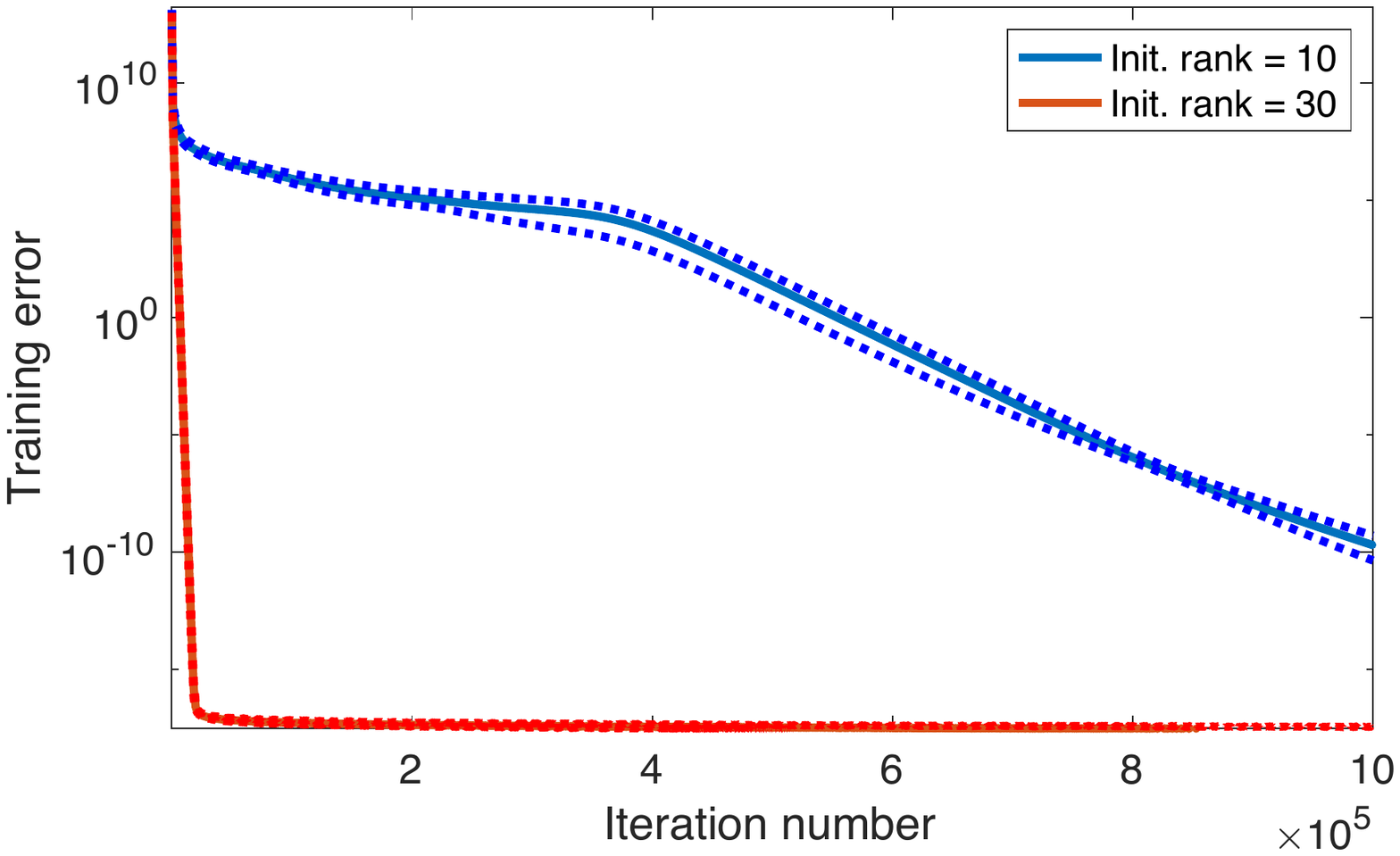}   
        \hfill
        \includegraphics[width=6.9cm,height=4.5cm]{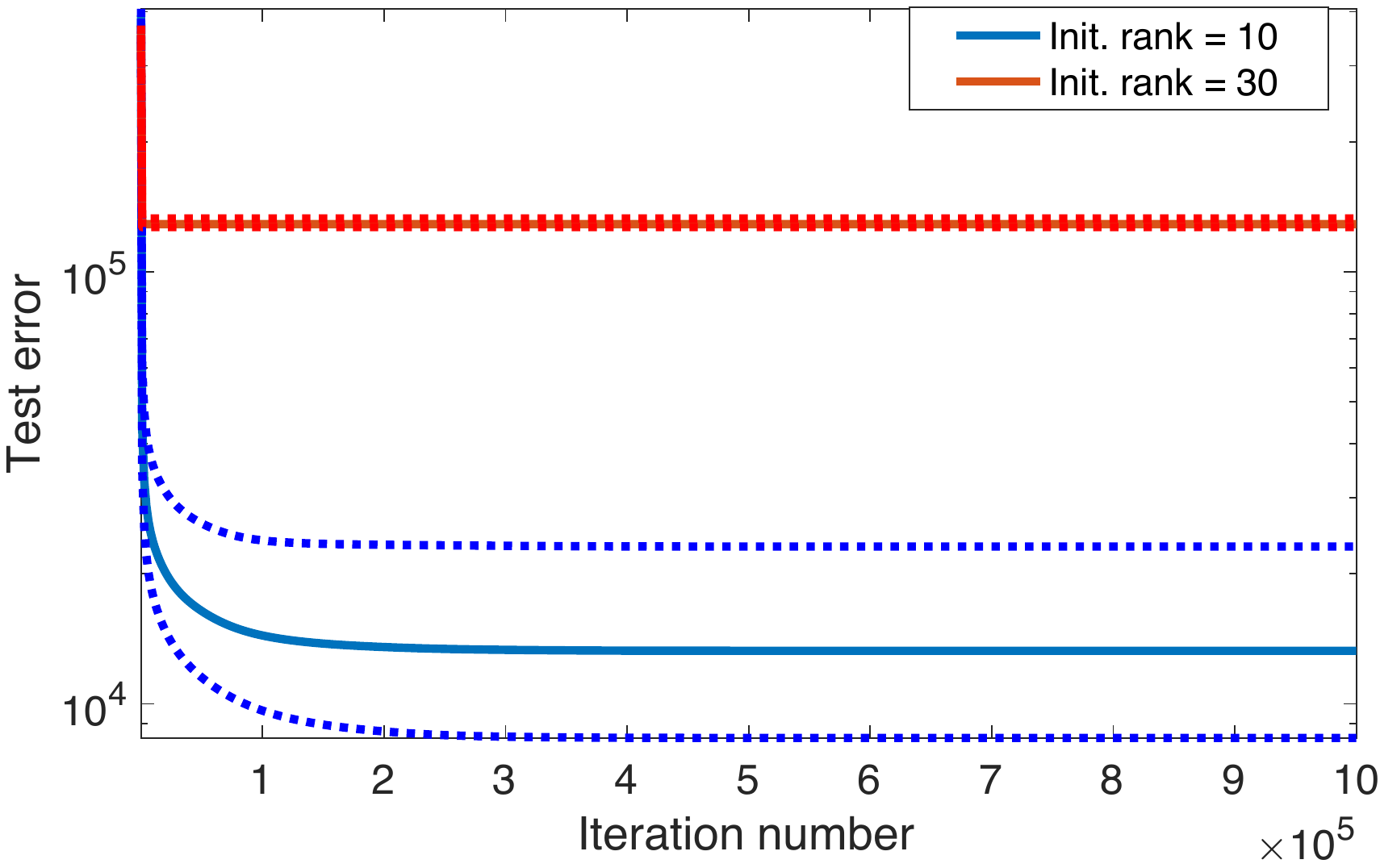}
        \caption{\label{fig:break} 
              This figure complements Figures~\ref{fig:lowNorm} and~\ref{fig:hiNorm}.
        With the setup explained in Example~\ref{ex:numEx}, this figure shows the training error (left) and test error (right)  of the gradient descent~\eqref{eq:gradDec}, when initialized much further from the origin $(\|U_0\|_\F=10^3)$. The solid and dotted lines again show the average over three trials  $\pm$ half of the standard deviation. 
        Note that the test error is large for both initialization ranks, i.e., the gradient descent~\eqref{eq:gradDec} is \emph{not} implicitly biased towards the planted model~$X^\n$.
        Nevertheless, the test error is smaller  when the initialization is low-rank. 
        % {One of our technical assumptions for successful implicit regularization is that the gradient descent~\eqref{eq:gradDec} must be initialized sufficiently near the origin, see~\ref{item:nearby} in Section~\ref{sec:contributions}.  This figure visualizes the necessity of this assumption. More specifically, with the same setup as in Example~\ref{ex:numEx}, this figure shows the training error (left) and test error (right)  of the gradient descent~\eqref{eq:gradDec}, when initialized very far from  the origin $(\|U_0\|_\F=10^3)$.
        % Note  that $U_0U_0^\top$ is far from the planted model~$X^\n$  because $\trace(U_0U_0^\top)\gg \trace(X^\n)=1$. 
        % The solid and dotted lines show the average over three trials  $\pm$ half of the standard deviation.
        % Note that the test error is large and the gradient descent~\eqref{eq:gradDec} is not implicitly biased towards the planted model~$X^\n$. }
        % Gradient descent consistently displays implicit regularization towards the  planted low-rank matrix, regardless of initialization norm, suggesting that it is the initialization rank rather, than its norm, which is chiefly responsible for the implicit regularization of the gradient descent~\eqref{eq:gradDec}. 
        }
    \end{figure}

 The numerical Example~\ref{ex:numEx} uncovers a new facet about the implicit {bias of  gradient flow~\eqref{eq:flowIntro}},
%  regularization, 
 i.e., its sensitivity to  the initialization rank.
 For instance,  when initialized {\emph{high-rank}}, the flow~\eqref{eq:flowIntro}  is in general {\emph{not}} implicitly biased towards the planted matrix $X^\n$, see Figure~\ref{fig:hiNorm}.
%  That is, we show here that  the implicit \note{bias} of the flow~\eqref{eq:flowIntro} also depends strongly on the {rank} of its initialization. 
%  {In particular, when initialized high-rank,  in general the flow~\eqref{eq:flowIntro} is \emph{not} implicitly biased towards the planted model $X^\n$, see Figure~\ref{fig:hiNorm}. 
 Towards better understanding this new limitation of implicit bias, we  partially characterize the implicit bias of the flow~\eqref{eq:flowIntro}, when  its initialization is low-rank. 
% Indeed, the numerical evidence suggests that  the flow~\eqref{eq:flowIntro}, with  a generic low-rank initialization, might be better   biased towards
% the low-rank planted model~$X^\n$, compared to a high-rank initialization. 
% More specifically, the theoretical contribution of our work is to explain this numerical observation, albeit in a limited setting.
Informally speaking, our main theoretical finding  is that: 
\begin{gather}
\textit{When initialized low-rank and with a sufficiently small training (or test) error, the gradient flow~\eqref{eq:flowIntro}} \nonumber\\
\textit{is implicitly biased towards the planted low-rank model } X^\n  \textit{, and  achieves zero test error.}
\label{eq:finding}
\end{gather}
After recalling~\eqref{eq:nearlyZero}, we observe that \eqref{eq:finding} provides an alternative to the  full-rank and nearly-zero initialization scheme of~\cite{gunasekar2017implicit,arora2019implicit,li2018algorithmic}. Note also that both of the assumptions in~\eqref{eq:finding}
loosely reflect the real limitations of implicit bias that we numerically identified in Example~\ref{ex:numEx}:
% \circled{1}~

First, when initialized high-rank, the flow~\eqref{eq:flowIntro}  is in general \emph{not} implicitly biased towards $X^\n$,  unless the initialization is near the origin as in~\eqref{eq:nearlyZero}.
% ~\circled{2}~
Second, if the initial test error is very large, then the initialization is  far from the origin (by triangle inequality). In turn, when initialized faraway, we saw in Figure~\ref{fig:break} that the flow~\eqref{eq:flowIntro} is in general \emph{not} implicitly biased towards $X^\n$. As we will see later, our theoretical finding in~\eqref{eq:finding} is also fundamentally different from the local refinement results in signal processing~\cite[Chapter 5]{chi2019nonconvex}.

To complement the theoretical findings in~\eqref{eq:finding}, we also propose a more  practical scheme to recover~$X^\n$:
% This new scheme can be seen as an alternative to the full-rank and almost-zero initialization heuristic which has dominated the literature~\cite{gunasekar2017implicit,arora2019implicit,li2018algorithmic}. 
Indeed, the gist of Example \ref{ex:numEx} and~\eqref{eq:finding} is that both high-rank and high-norm initializations are in general detrimental to implicit bias. This observation later leads us to the following heuristic:
% which replaces  the ``sufficiently small test error'' requirement in~\eqref{eq:finding} with a more practical guideline:  
\begin{gather}
\textit{To recover the planted matrix $X^\n$, initiate the flow \eqref{eq:flowIntro}} \textit{ with a random matrix.} \nonumber\\
\textit{If the training error} 
% \|\A(U(t)U(t)^\top)-b\|_2^2 
\textit{ is reducing too rapidly,}  \nonumber\\
\textit{then restart the training after decreasing the initialization rank and norm.}
% \nonumber\\
% \textit{restart the training with a new random initialization.}
\label{eq:initHeuristic}
\end{gather}
% We will make \eqref{eq:initHeuristic}  precise later on. 
We may again think of \eqref{eq:initHeuristic} as an alternative that complements the full-rank and almost-zero initialization scheme that is predominant in the literature of matrix sensing~\cite{gunasekar2017implicit,arora2019implicit,li2018algorithmic}, see~\eqref{eq:nearlyZero}. 
{Let us next present a more detailed, but still simplified,  version of our main theoretical result in~\eqref{eq:finding}, specialized to the case where $\A$ is a generic linear operator. 
%  After a formal problem statement in Section~\ref{sec:setup} and literature review in Section~\ref{sec:relatedWork}, we first provide a geometric perspective  {of implicit regularization} in Section~\ref{sec:nonConvexGeom}, and then prove the main result of this work in Section~\ref{sec:mainResults}, summarized below informally.

\begin{thm}[\textsc{Main result,  simplified}]\label{corr:mainSimple}
Consider the model $X^\n\in\R^{d\times d}$ and fix an integer~$m$. For every $i\le m$, populate the upper triangular entries of the symmetric matrix $A_i\in \R^{d\times d}$ with independent Gaussian random variables that have zero mean and unit variance. In doing so, we obtain a generic sensing operator $\A:\R^{d\times d}\rightarrow\R^m$ that maps $X$ to $[\langle A_1,X\rangle, \cdots, \langle A_m, X\rangle]^\top$. 
Let $b:=\A(X^\n)$ denote the training data.
% Minimizing $\|\A(UU^\top)-b\|_2^2$ with  the
% fix an integer $p\ge \rank(X^\n)$. 
When initialized at~$U_0\in \R^{d\times d}$, the limit point of the gradient flow \eqref{eq:flowIntro} exists and achieves zero test error, provided that 
% has a limit point $\ol{U}$ that
% achieves zero test error $(\ol{U}\,\ol{U}^\top= X^\n)$, provided that: 
\begin{enumerate}[leftmargin=*,label={(\roman*)}]
    \item\label{item:iso} $m=\Omega(\rank(U_0) d \log d)$, i.e., $m \gtrsim \rank(U_0) d \log d$, where $\gtrsim$ suppresses the constants.
    \item\label{item:initRank} $\rank(U_0) \ge \rank(X^\n)$. 
    \item\label{item:nearby} $\|U_0\|\le \sqrt{\|X^\n\|}$, where $\|\cdot\|$ denotes the spectral norm.
    \item \label{item:nearby2}    \editae{The initial training (or test) error is  not too large}. This will be made precise later. 
    
    % \item\label{item:manifold} The set $\{U:\A(UU^\top)=b\}$ is a smooth manifold.
\end{enumerate}
\end{thm}
% The above result is detailed in Theorem~\ref{thm:main}. 
For clarity and insight, it is not uncommon to study the gradient flow \eqref{eq:flowIntro} as a proxy for its discretization~\eqref{eq:gradDec}, e.g., see~\cite{arora2019implicit,eftekhari2020training}.  
Let us now informally justify the  assumptions of Theorem~\ref{corr:mainSimple}.
% leaving the details for later.
% Section~\ref{sec:mainResults}.
\begin{itemize}[leftmargin=*,wide]
    \item  With overwhelming probability, \ref{item:iso} ensures that the linear operator $\A$ is injective, when restricted to the set of all matrices with rank at most $ \rank(U_0)$~\cite{geyer2020low,foucart2019iterative}. 
    % but  a direct comparison with these works is difficult because the initialization regime  is different here. 
    This restricted injectivity property (RIP)  is necessary because, otherwise, the planted model~$X^\n$ might not be identifiable from the training data~$b=\A(X^\n)$. For instance, in the absence of \ref{item:iso}, we observe from \ref{item:initRank} that $U_0U_0^\top$ and $X^\n$ might be indistinguishable and both mapped by the operator $\A$ to the same training data $b$ !

    {Crucially, in order for~{\ref{item:iso}} to hold in practice, the initialization rank   cannot be too large. More specifically, $m$ is in practice often limited by our sampling budget. 
    Therefore, in order to enforce {\ref{item:iso}}, we must in turn select $\rank(U_0)$ to be relatively small, i.e., $\rank(U_0)=O(m/(d\log d))$.  
    In this way, Theorem~\ref{corr:mainSimple} reflects the true limitations of implicit bias that we numerically identified in Example~\ref{ex:numEx}. Indeed, recall that the flow~\eqref{eq:flowIntro}  is in general \emph{not} implicitly biased towards $X^\n$, when its initialization is high-rank, see Figure~\ref{fig:hiNorm}. }
    Lastly, the RIP in~\ref{item:iso} is in the same vein as~\cite{li2018algorithmic,geyer2020low}. 
    
% For example,~\cite{li2018algorithmic} studies a full-rank initialization near the origin, whereas  
    \item {\ref{item:initRank}} is  weaker than $\rank(U_0)=d$  in~\cite{gunasekar2017implicit,li2018algorithmic}.  In this sense, Theorem~\ref{corr:mainSimple} is the first result to investigate the implicit bias of the gradient flow \eqref{eq:flowIntro} in the more general setting where $\rank(U_0)\ge \rank(X^\n)$. 
    
    \item {\ref{item:nearby}}  reflects the  limitations of implicit bias in the following sense: As we observed in the numerical Example~\ref{ex:numEx}, the flow~\eqref{eq:flowIntro} is in general \emph{not}  implicitly biased towards the planted model $X^\n$, when initialized far from the origin, see Figures~\ref{fig:hiNorm} and~\ref{fig:break}.  
    % the implicit bias towards the planted model $X^\n$ gradually disappears as the flow~\eqref{eq:flowIntro}  is initialized further and further from the origin. 
    At the same time, recall also from~\eqref{eq:nearlyZero}  that~\cite{gunasekar2017implicit,arora2019implicit,li2018algorithmic} only study an almost-zero initialization for the flow~\eqref{eq:flowIntro}. Compared to these works, Theorem~\ref{corr:mainSimple} is the first result to investigate the implicit bias of~\eqref{eq:flowIntro} beyond the vanishing initialization norm in \eqref{eq:nearlyZero}. 
    % For instance, with the \emph{same} planted model~$X^\n$ and sensing operator~$\A$ and initialization ranks as in Example~\ref{ex:numEx},  Figure~\ref{fig:break} shows no implicit bias towards $X^\n$ (large test error) when the gradient descent~\eqref{eq:gradDec} is initialized \editae{very} far from the origin. 

    \item Likewise, {\ref{item:nearby2}} also partially mirrors the limitations of implicit bias in the following sense: It follows from the triangle inequality that an initialization $U_0$ with a large test error  ($\|U_0U_0^\top-X^\n\|_\F\gg \|X^\n\|_{\F}$) is also far from the origin ($\|U_0\|_\F\gg 0$).  In turn, as we saw earlier in Figures~\ref{fig:hiNorm} and~\ref{fig:break}, in general the gradient flow~\eqref{eq:flowIntro} is \emph{not} implicitly biased towards $X^\n$, when initialized faraway from the origin. 
    % again leads to an absence of implicit bias (large test error), as explained above and in Figure~\ref{fig:break}.  
    
    We emphasize that Theorem~\ref{corr:mainSimple} is an example of a ``capture theorem'' within the literature of nonconvex optimization~\cite{boumal2016non,sahin2019inexact,li2018calculus,attouch2013convergence}. As is standard, our capture theorem  predicates on an initialization within a specific ``capture neighborhood'' of the set of matrices with zero training error, i.e., our result predicates on  an initialization  within a certain neighborhood of the set $\{U:\A(UU^\top)=b\}$. 
    
    Such capture theorems are fundamentally different from the local refinement results, within the signal processing literature. The latter group of results often rely on  local strong convexity in a very small neighborhood of a matrix with zero test error~\cite[Chapter 5]{chi2019nonconvex}. {In contrast, our capture neighborhood contains \emph{all} matrices with zero training error (which of course includes all matrices with zero test error). We can visualize this capture neighborhood as a ``tube'' $\{U:\|\A(UU^\top-\underline{U}\,\underline{U}^\top)\|_2\le \rho\}$ } rather than a small Euclidean ball centered at a matrix $\underline{U}$ with zero test error ($\underline{U}\,\underline{U} ^\top=X^\n$). 
    % Here, $a$ is a geometric attribute of the problem, which will be made precise later. 
    % in which~\ref{item:nearby} is violated and the gradient flow~\eqref{eq:flowIntro} successfully interpolates the training data in limit but fails to recover the planted model~$X^\n$, regardless of initialization rank.
    % \item Lastly, the manifold assumption in~\emph{\ref{item:manifold}}   is  minimal in the sense that it corresponds to the weakest sufficient conditions under which there is any hope for the flow~\eqref{eq:flowIntro} to efficiently find a limit point with zero training error. See also~\cite{boumal2020deterministic,sahin2019inexact} for precedents of~\emph{\ref{item:manifold}} within the nonconvex optimization literature.
\end{itemize}
%   In this figure, the setup is identical to Example~\ref{ex:numEx}, but the initialization norm is~$\|U_0\|_\F = 10^3$. Consequently,~$U_0U_0^\top$ is far $X^\n$ because $\trace(U_0U_0^\top)=10^6\gg 1=\trace(X^\n)$. 
To summarize, this work is a small step towards  better understanding the limitations of implicit bias in matrix sensing. Here, we identify and partially quantify  the sensitivity of implicit bias 
% of \eqref{eq:flowIntro}
to the initialization rank. 
% As we discussed earlier,
% ~\emph{\ref{item:nearby2}} in Theorem~\ref{corr:mainSimple} is difficult to verify in practice. 
% Consequently, 
% Theorem \ref{corr:mainSimple} is primarily a theoretical result which sheds light on the theoretical aspects of the implicit bias.
% and
% cannot be used as a practical initialization scheme for the flow~\eqref{eq:flowIntro}.
% This  shortcoming  is ubiquitous in ``capture theorems'' within the literature of nonconvex optimization~\cite{sahin2019inexact}. 
% To supplement our theoretical findings, 
% Theorem~\ref{corr:mainSimple}, 
We also propose an adaptive heuristic for matrix sensing which complements the predominant high-rank and near-zero initialization scheme~\cite{gunasekar2017implicit,arora2019implicit,li2018algorithmic}.  Beyond matrix sensing, a near-zero initialization often leads to vanishing gradients in deep learning~\cite{hochreiter2001gradient}. Alternatives to \eqref{eq:nearlyZero}, such as our~\eqref{eq:initHeuristic},  might in the future provide valuable insights for more difficult learning problems. 
% Loosely speaking, this scheme restarts the training with a new low-rank initialization if 

} 
% The practical aspects of the initialization, alongside many other features are carefully discussed in Section~\ref{sec:mainResults}. In particular, 
% Nevertheless we will also discuss the practical aspects in Remark~\ref{rem:practicalInit}.
% Let us now turn to the details. 

\vspace{-5pt}
\section{Objective and Assumptions}\label{sec:setup}
\vspace{-5pt}

%  {To begin, this section formally states our problem, assumptions and objective.}
 Here, we set the stage for our detailed main result. 
For $\xi>0$, consider a   model  {$X^\n\in \R^{d\times d}$} such that 
\begin{align}
    0 \preceq X^\n \prec \xi^2 I_d, 
    % \,\,\, X^\n \in \R^{d\times d},
    \tag{model}
    \label{eq:model}
\end{align}
where
$I_d\in \R^{d\times d}$ is the identity matrix, and $A\preceq B$ means that $B-A$ is a positive semi-definite~(PSD) matrix.  {Likewise, $A\prec B$ means that $B-A$ is a positive definite matrix.} {As suggested by~\eqref{eq:model},}  we  limit ourselves to PSD  matrices throughout, similar to~\cite{gunasekar2017implicit,li2018algorithmic}.
Note also that~\eqref{eq:model} conveniently assumes \emph{a priori} knowledge of ({an} estimate of)~$\|X^\n\|$, where~$\|\cdot\|$ stands for the spectral norm. To gauge the complexity of a model, we will rely on rank and effective rank, defined below.

\begin{figure}
        \centering
        \includegraphics[width=8cm,height=6cm]{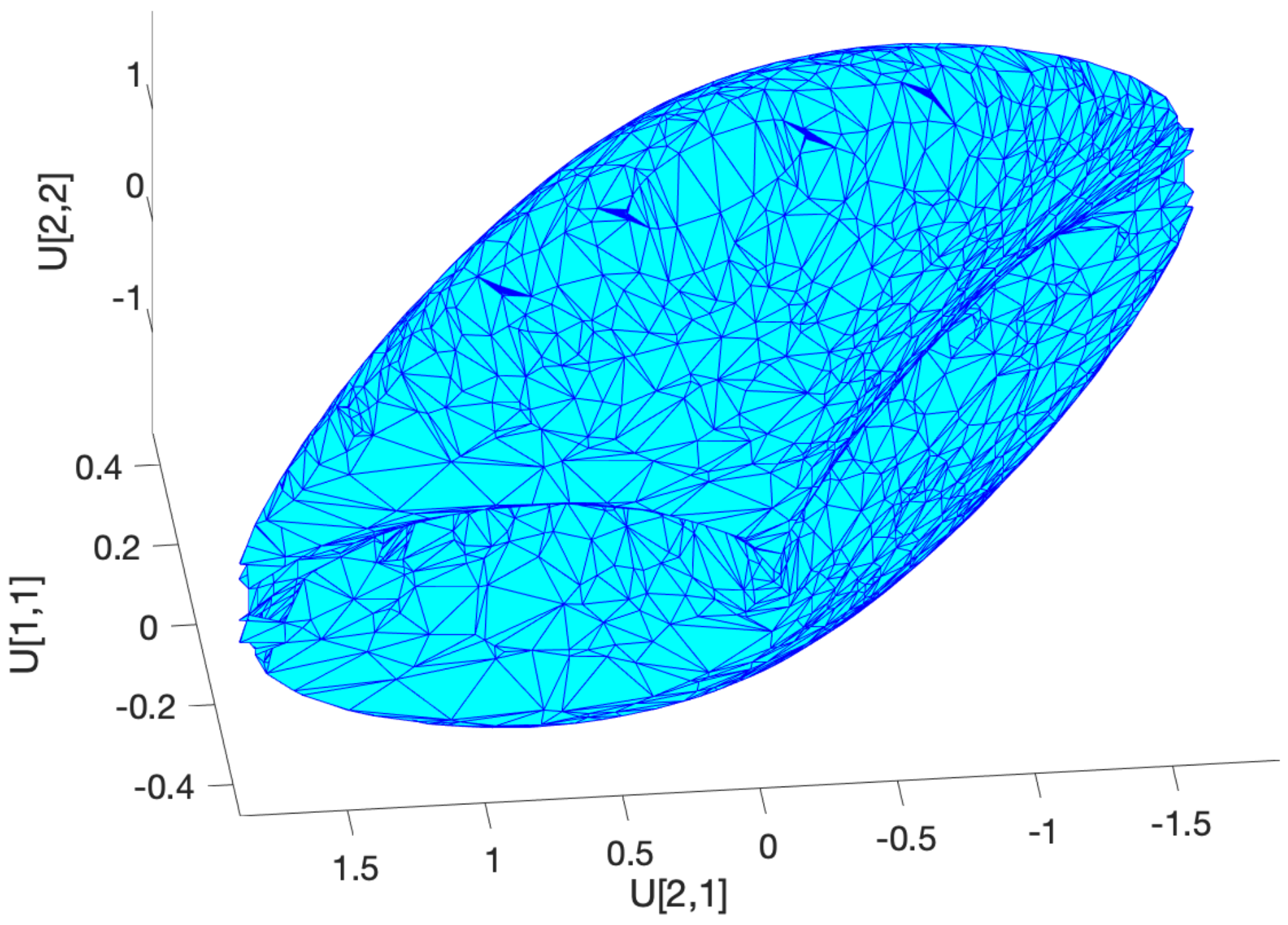}   
        \caption{This figure visualizes parts of the set~$\M_b$ for a toy example.
        Defined in~\eqref{eq:manifold}, recall that~$\M_b$ is the set of all matrices with zero training error and bounded norm. Above,~$d=p=m=2$ and~$X^\natural\in\R^{d\times d}$ in~\eqref{eq:model} is a random matrix of rank~$r=1$. The linear operator~$\A:\R^{d\times d}\rightarrow\R^m$ takes~$UU^\top\in \R^{d\times d}$ to~$\A(UU^\top)=[\|U[1,:]\|_{2}^2 \,\, \langle A, UU^\top \rangle]^\top\in \R^m$ for a random matrix~$A\in\R^{2\times 2}$. Here,~$U[1,:]$ is the first row of~$U$ in MATLAB notation. Even though~$\M_b$ is a subset of~$\R^{d\times p}$, the particular structure of~$\A$ in this toy example allows us to visualize this set. More specifically, for every~$U\in\M_b$, the constraints~$\A(UU^\top)=b$ in~\eqref{eq:manifold} determine the $\ell_2$-norm of the first row of~$U$. Consequently, above we plot  only three entries of the matrix~$U$, namely,~$U[1,1]$,~$U[2,1]$ and~$U[2,2]$ in MATLAB notation. The three-dimensional mesh has been created with help from~\cite{mesh}.
        }
        \label{fig:vis}
    \end{figure}

\begin{defn}[Effective rank]\label{defn:effRank}
The effective rank of a PSD matrix $X$, denoted throughout by $\effrank(X)$, is the smallest integer $r$ such that  there exists a PSD matrix $X_r$, of rank at most~$r$, that is infinitesimally close to $X$.  {In particular},  $\|X-X_r\|_\F \le \epsilon \|X\|_\F $ for an infinitesimal $\epsilon$. The effective rank and (standard) rank are related as   $\effrank \le \rank$.
\end{defn}
The subtle distinction between $\rank$ and $\effrank$ is only for technical correctness, which the reader may 
% safely 
ignore in a first reading. 
The  {closely} related  {concepts of numerical or approximate rank and}  border rank  {are} discussed in~ {\cite[Page 275]{golub2013matrix} and~}\cite[Section 3.3]{kolda2009tensor}.
% respectively.  
% {respectively.}} 
%  We also make the mild assumption that $\xi$ is known \emph{a priori}, for the technical convenience of compactness throughout the analysis.
For symmetric matrices $\{A_i\}_{i=1}^m\subset \R^{d\times d}$, consider  the linear operator $\A:\R^{d\times d}\rightarrow \R^m$ and the vector~$b\in \R^m$, defined as 
\begin{align}
    \A(X) := [\langle A_1, X\rangle, \cdots, \langle A_m, X\rangle]^\top,
    \qquad 
    b := \A(X^\n).
    \tag{sense}
    \label{eq:measOp}
\end{align}
 {In the context of matrix sensing~\cite{davenport2016overview}, we may think of $\A$ and $b$ as the sensing operator and the training data, respectively.}
% and let 
% \begin{align}
%     b := \A(X^\n_l),
%     \label{eq:measOp}
% \end{align}
% where $X^\n_l$  is infinitesimally close to $X^\n$ and satisfies $\rank(X^\n_l) = \effrank(X^\n)$.
% In linear inverse problems,  $\A$ and $b$ are commonly referred to as the measurement operator and the measurements vector, respectively. 
% {We are now in position to state our objective.}
% \begin{tcolorbox}[width=1\textwidth,colback={gray!40},
% title={With rounded corners},
% colbacktitle=yellow,
% coltitle=blue
% ]    
% \begin{objective}\label{o:objective} 
Given the linear
operator $\A$ and the 
 {training data}
$b$  in \eqref{eq:measOp}, the objective of this work is to recover $X^\n$ in~\eqref{eq:model}, up to an infinitesimal error.
% \end{objective}
% \end{tcolorbox}  
% The rest of this section delineates our assumptions to recover the true model $X^\n$.
 {Towards that objective, for an integer $p\ge \rank(X^\n)$, it is   convenient to} define the maps 
  $g:\R^{d\times p}\rightarrow\R^m$ and
 $G:\R^{d\times p}\rightarrow\R$ as 
\begin{align}
g(U):= \frac{1}{2}( \A(UU^\top) - b),
\qquad 
      G(U) :=
       \frac{1}{8}\|\A(UU^\top) - b\|_2^2.
    \tag{training error}
    \label{eq:fgDefn}
\end{align}
 {In words, $G$ is the (scaled) training error associated with the model~$UU^\top$, i.e., $G$ gauges the discrepancy between the sensed vector $\A(UU^\top)$ and the training data~$b$. In particular,~$G(U)=0$ if and only if  the training error is zero, i.e., $G(U)=0$ iff $\A(UU^\top)=b$. Note also that $G$ above   might be a nonconvex function of~$U$.}
% For an integer $p\ge \rank(X^\n)$, 
Next, consider the set 
\begin{align}
    {\M_b} & := \{ U: \A(U U^\top) = b,\, \|U\|\le  \xi \}= \{U: G(U) = 0,\, \|U\|\le   \xi \} \subset \R^{d\times p},
     \label{eq:manifold} \tag{manifold}
\end{align}
%  {As we will see shortly, the set $\M_b$ and its geometry plays a key role in this work. For now,
% we only note} 
and note that every matrix $U\in \M_b$ corresponds to a model $UU^\top\in \R^{d\times d}$ with zero training error ($\A(UU^\top)=b$).  {Put differently, $\M_b$ is the set of all matrices, like~$U$, that have zero training error ($G(U)=0$) and satisfy $\|U\|\le \xi$, as visualized in Figure~\ref{fig:vis}}. 
% Recalling  {the definition of $g$ in}~\eqref{eq:fgDefn}, 
% For future reference, note  {also} that the {(total)} derivative of~$g(U)$ at $U$ is the linear operator  $\Der g(U):\R^{d\times p} \rightarrow\R^m$, defined as
% \begin{align}
%     \Der g(U)[\D] := \A(\D U^\top).
%     \qquad \text{(see \eqref{eq:fgDefn})}
%     \label{eq:defnC}
% \end{align}
% and its adjoint  {operator} is denoted by $(\Der g(U))^*$.
% specified as
% \begin{equation}
%     (\Der g(U))^* [\d] := \A^*(\d)\cdot U = \sum_{i=1}^m \d_i A_i U,
%         \label{eq:adjO}
% \end{equation}
% where $\A^*$ is the adjoint of the linear operator $\A$ in~\eqref{eq:measOp} and $\d_i$ is the $i^{\text{th}}$ entry of the vector~$\d\in \R^m$. 
 To recover~$X^\n$ in~\eqref{eq:model} from~$\A$ and $b$ in~\eqref{eq:measOp}, we will use gradient flow to minimize the~\eqref{eq:fgDefn}. More formally, consider 
%  the optimization problem 
 \begin{align}
     \min_{U\in \R^{d\times p}}\, G(U) =  {\frac{1}{8}\|\A(UU^\top)-b\|_2^2},
    %  \,\, \text{subject to}\,\, \|U\|\le \xi,
    \tag{ERM}
     \label{eq:feas}
 \end{align}
  {where ERM stands for empirical risk minimization~\cite{shalev2014understanding}, and above we used the definition of~$G$ in}~\eqref{eq:fgDefn}. 
%   {the} problem~\eqref{eq:feas}  {provably} has no spurious second-order stationary points~(SOSPs). i.e.,  {every} SOSP of problem~\eqref{eq:feas} is also a global minimizer of problem~\eqref{eq:feas} with  {the} optimal value of zero, see~\cite[Theorem~33]{chi2019nonconvex}.  {The benign landscape of the problem~\eqref{eq:feas} also means that  problem~\eqref{eq:feas} is amenable to fast optimization.}
To recover~$X^\n$, we apply the gradient flow to~\eqref{eq:feas} and obtain
 {\begin{align}
    \dot{U}(t) &  =   - \nabla G(U(t))   = - \frac{1}{2}\A^*\left( \A(U(t)U(t)^\top) - b\right) U(t), \qquad U(0) = U_0\in \R^{d\times p}.
    \tag{flow}
    \label{eq:flowG}
\end{align}}where $\A^*:\R^m\rightarrow\R^{d\times d}$ denotes the adjoint of the operator~$\A$ in~\eqref{eq:measOp}.

A crucial observation is the following:
Under our key Assumption~\ref{assumption:key} below, 
% {we see by counting the degrees of freedom} that the  operator $\A$ in~\eqref{eq:measOp}  \emph{cannot} satisfy the $2p$-RIP. As 
% verified in the supplementary material, this observation implies  
 the set~$\M_b$ in~\eqref{eq:manifold}   {may} contain infinitely many  {distinct matrices with zero training error but nonzero test error}, see Example~\ref{ex:numEx}. 
%  (See the supplementary material for this claim.)  
 {That is, there might exist infinitely many matrices like $U\in \R^{d\times p}$ that satisfy $\A(UU^\top) = b$, but~$UU^\top$
 is \emph{far} from  {the planted model}~$X^\n$.}  
  {Put differently, under Assumption~\ref{assumption:key}, problem~\eqref{eq:feas} might have infinitely many global minimizers, like $U$, that reach the optimal value zero of problem~\eqref{eq:feas},
%  (i.e., $G(U)=0$), 
 but~$UU^\top$ is far from the planted model~$X^\n$. 
%  See Figure~\ref{fig:hiNorm}, for instance. 
 Indeed, note that problem~\eqref{eq:feas} lacks any explicit bias towards $X^\n$ (such as regularization with the trace norm~\cite{davenport2016overview}).
%  Nevertheless, when is the~\eqref{eq:flowG}  implicitly biased towards the planted model $X^\n$? 
%  Asked differently, is the~\eqref{eq:flowG} implicitly biased towards a limit point $\ol{U}\in \R^{d\times p}$ such that $\ol{U} \,\ol{U}^\top = X^\n$? 
Nevertheless, does the \eqref{eq:flowG} recover $X^\n$? We answer this question in Section~\ref{sec:mainResults} and then compare our main result with the existing literature in Section~\ref{sec:relatedWork}.

 }

\vspace{-5pt}
\section{Main Result}\label{sec:mainResults}
\vspace{-5pt}

Our main result below posits that, when initialized properly, {the}~\eqref{eq:flowG} converges  to a limit point 
% $\ol{U}\in \R^{d\times p}$  such that $\ol{U}\cdot \ol{U}^\top $ is infinitesimally close to $X^\n$,
 {with zero test error,}
thus 
successfully recovering the planted model $X^\n$. We first collect our assumptions.

  {
% A simple but important corollary below replaces $\dist(U_0,\M_b)$ in Theorem~\ref{thm:main} with $G(U_0)$ in~\eqref{eq:feas}. The corollary below is the main result of this work and will be followed by several remarks. 

\begin{assumption}\label{assumption:key}
 {Consider the framework specified by~\eqref{eq:model} and~\eqref{eq:measOp}.}
For an integer $p$, consider also  the~\eqref{eq:flowG},  initialized at $U_0\in \R^{d\times p}$. 
 {
\begin{enumerate}[leftmargin=*,label={(\roman*)},wide]
\item (Recoverable). \label{assumption:infoLevel} Assume that   $m/d \le \rank(X^\n) \le p$.
% \item ({Over-parametrized}).\label{assumption:badRegime}  {$m<pd$ if $p<d/2$ and $ m < d(d+1)/2$, otherwise.}
\item \label{assumption:man} (Manifold). Assume that $\rank(\Delta \rightarrow \A(\Delta U^\top)) = m$,
% \le  \rank(X^\n) d$,
% such that
% \begin{align}
    % \rank(\Delta \rightarrow \A(\Delta U^\top)) = m, 
    % \label{eq:fullRankAll}
% \end{align}
for every $U$ that belongs to an open neighborhood of the set $\M_b$ in~\eqref{eq:manifold}.
Equivalently, assume that {the matrices}~$\{A_i U\}_{i=1}^m$ span an $m$-dimensional subspace of $\R^{d\times p}$, for every $U$ in an open neighborhood of~$\M_b$. 
% If $\s_m(\cdot)$ denotes the $m$th largest singular value of an operator, this assumption can be formulated as
% \begin{equation}
%     \s_m(\M_b) := \min\{ \s_m(\Delta \rightarrow \A(\Delta U^\top) ) : U\in \M_b \} >0. 
%     \label{eq:sManifoldAssump}
% \end{equation}

\item (Restricted injectivity).\label{assumption:iso} Assume that the linear operator $\A$ in~\eqref{eq:measOp} satisfies the $(2\rank(U_0))$-restricted injectivity property, abbreviated as  $(2\rank(U_0))$-RIP. More specifically, we assume that $\A(X)=0\Longrightarrow X=0$,
% , up to an infinitesimally small error, 
% there exist scalars $0< \a \le \b$ such that 
% $\a \|X\|_\F \le \|\A (X) \|_2 \le \b \|X\|_\F$,
for every PSD matrix $X$ with $\rank(X)\le 2\rank(U_0)$. We let $\alpha$ denote the corresponding isometry constant, i.e., $\alpha :=\min\{\|\A(X)\|_{\F}:\|X\|_{\F}=1,\, \rank(X)\le 2\rank(U_0) \}$.

    % \item\label{item:nearbyCorr} $\|U_0\|\le \xi$, 
    \item \label{item:initRankLarge} Assume that $\rank(U_0)\ge \effrank(X^\n)$,
    \item\label{item:howSmallInitV2Corr} Recall $\xi$ from \eqref{eq:model}. Assume that  $U_0$  satisfies $\|U_0\|\le \xi$ and  \underline{any} of the  inequalities
    \begin{subequations}\label{eq:3Sub}
    \begin{align}
     & \|\A(U_0U_0^\top) - b\|_2  <
     \frac{ \s_m(\M_b)\s_{\min}(\M_b) \alpha}{4 \|\A\|}
    ,
     \qquad \text{(initial training error)}
     \label{eq:a}
     \\
     & \| U_0U_0^\top - X^\n\|_\F <
     \frac{\s_m(\M_b) \s_{\min}(\M_b) \alpha}{4\|\A\|^2}
     ,
     \qquad \text{(initial test error)}
     \label{eq:b}
     \\
      & \dist(U_0,\M_b)  <\frac{\s_m(\M_b)}{4\|\A\|} \min\l( 1,\frac{ \s_{\min}(\M_{b})\a }{4 \xi \|\A\|} \r).
      \qquad \text{(initial distance to $\M_b$)}
    \label{eq:howSmallInitV2Corr}
    \end{align}
    \end{subequations}
    Above, $\s_m(\M_b):=\min\{\s_m(\Delta\rightarrow\A(\Delta U^\top)): U\in \M_b \}$, where $\s_m(\cdot)$  returns the $m$th largest singular value of an operator. 
    % was defined in~\eqref{eq:sManifoldAssump} and 
    Also, $\s_{\min}(\M_b):= \min\{ \s_{\min}(U): U\in \M_b \}$, where $\s_{\min}(\cdot)$ returns the smallest nonzero singular value of an operator.  
    % ~$\rho_0:=\min\{\sigma_m(\Der g(U)): U\in \M_b\}/(2\|\A\|)$, and 
    Lastly, $\alpha$ is the isometry constant and the operator norm of the linear operator~$\A$ in~\eqref{eq:measOp}, respectively. 
    % Lastly, $\s_{\min}(\M_{b,\rho_0})$ is the smallest nonzero singular value in the $\rho_0$-neighborhood of $\M_b$, i.e.,  $\M_{b,\rho_0}:=\{ U: \exists U'\in \M_b,\, \|U'-U\|_{\F}\le \rho_0\}$. 
    % both defined precisely in Appendix~\ref{sec:proofThm}. 
    % and~$\|\A\|$  is the operator norm of~$\A$.
    
\end{enumerate}}

\end{assumption}

\begin{thm}[\textsc{Implicit bias, main result}]\label{thm:main} 
%  Suppose  {also} that Assumption~\ref{assumption:iso}  holds with {$r\ge 2\cdot \rank(U_0)$}. { {Finally,} 
Consider the framework specified by~\eqref{eq:model} and~\eqref{eq:measOp}.
Suppose that Assumption~\ref{assumption:key} is fulfilled.
Then the~\eqref{eq:flowG} converges to a limit point $\ol{U}\in\R^{d\times p}$ such that~$\ol{U}\, \ol{U}^\top$ is infinitesimally close to~$X^\n$ in~\eqref{eq:model}.
\end{thm}

We refer to Section~\ref{sec:contributions} for  insights, and to Section \ref{sec:relatedWork} for comparison  with the  literature. 
% As discussed in Section~\ref{sec:contributions}, Theorem~\ref{thm:main} is designed to  reflect the limitations of implicit bias of the~\eqref{eq:flowG} in matrix sensing.  We refer to Section~\ref{sec:relatedWork} for a detailed comparison of Theorem~\ref{thm:main} with the related literature.
As discussed in Remark \ref{rem:initRank}, the right-hand sides of  \eqref{eq:3Sub} are often difficult to estimate in practice. Indeed, Theorem~\ref{thm:main} primarily provides theoretical insights and will later be complemented with a practical algorithm.

% Another remark below explains the derivation of  Theorem~\ref{corr:mainSimple} as a corollary of Theorem \ref{thm:main}. 
% in Section~\ref{sec:contributions} can be derived as a corollary to Theorem~\ref{thm:main}. 

% Among other topics, the following remarks also  explain how to derive Theorem~\ref{corr:mainSimple} in the introduction as a corollary of Theorem~\ref{thm:main}. 
%  {Unlike Theorem~\ref{thm:srebro} from~\cite{li2018algorithmic}, note that Theorem~\ref{thm:main} is not limited to the case $p=d$ and also guarantees the convergence of the~\eqref{eq:flowG}.
% % addresses both of the concerns raised in Remark~\ref{rem:initSrebr}. More specifically, unlike  Theorem~\ref{thm:srebro},~\eqref{eq:flowG} converges and also need not necessarily be initialized near the origin.
%   More importantly, Theorem~\ref{thm:main} is the first theoretical evidence that partially explains the role of initialization rank in the implicit bias of gradient flow, a role which we numerically identified in Section~\ref{sec:intro}. 

%   Informally, a generic linear operator~$\A$ satisfies the RIP with high probability and the~\eqref{eq:flowG} successfully learns the planted model $X^\n$ from the training data $b$, whenever the initialization rank is sufficiently small and within the ``capture neighborhood''. 

\begin{remark}[\textsc{Proof outline}]
 At a high level, we follow a two-pronged approach: \circled{1}~We  establish that the~\eqref{eq:flowG} solves problem~\eqref{eq:feas} in the limit of \editae{$t\rightarrow\infty$}. That is, we  show that the limit point of the~\eqref{eq:flowG} exists and has zero training error.~\circled{2}~We also show that rank does not increase along the~\eqref{eq:flowG}, which allows us to prove that the test error of the~\eqref{eq:flowG} also vanishes in limit and that we  recover the planted model~$X^\n$, as desired. The details are deferred to the appendix. 
%  supplementary material. 

\end{remark}

\begin{remark}[\textsc{Recoverable}]\label{rem:recoverable} The inequality $\rank(X^\n)\le p $ in Assumption~\ref{assumption:key}\ref{assumption:infoLevel} is a trivial assumption which ensures that  {the set}~$\M_b$ in~\eqref{eq:manifold} is not empty. In particular, Assumption~\ref{assumption:key}\ref{assumption:infoLevel} ensures that there exists \underline{a} matrix $U\in \R^{d\times p}$ such that $\A(UU^\top) = b$. That is, under Assumption~\ref{assumption:key}\ref{assumption:infoLevel}, the optimal value of~\eqref{eq:feas} is zero  and there exists \underline{a} matrix $U\in \R^{d\times p}$ with zero~\eqref{eq:fgDefn}. Note that \cite{gunasekar2017implicit,arora2019implicit,li2018algorithmic} set $p=d$ and this condition
% Assumption~\ref{assumption:key}\ref{assumption:infoLevel} 
is thus met immediately.

On the other hand, the inequality $m/d\le \rank(X^\n)$ in  Assumption~\ref{assumption:key}\ref{assumption:infoLevel} is a mild technical assumption for the proofs. If not met at first, this inequality can be easily enforced by infinitesimally perturbing $X^\n$, as long as $m\le pd$.
Note that this perturbation does not change $\effrank(X^\n)$, see Definition~\ref{defn:effRank}. Only $\effrank(X^\n)$ is of statistical significance, and not $\rank(X^\n)$. Lastly, the simpler under-parametrized regime of $m>pd$ is not of interest here and we refer the reader to the survey \cite{chi2019nonconvex}.
\end{remark}

% \begin{remark}[\textsc{Over-parametrized}]
% Assumption~\ref{assumption:key}\ref{assumption:badRegime} reflects the limited sampling budget that is common in matrix sensing. Viewed differently, Assumption~\ref{assumption:key}\ref{assumption:badRegime} corresponds to the over-parametrized regime that is of interest to us in this work. In this over-parametrized regime, $p$ is relatively large compared to $m/d$. In particular, \cite{gunasekar2017implicit,arora2019implicit,li2018algorithmic} take $p=d$.
% Our results are in fact valid even if Assumption~\ref{assumption:key}\ref{assumption:badRegime} is violated but we chose to impose this assumption here because matrix sensing in the under-parametrized regime is considerably simpler, as detailed in the  supplementary material.
% \end{remark}

\begin{remark}[\textsc{Manifold}]\label{rem:manifold}
Assumption~\ref{assumption:key}\ref{assumption:man} 
% appears also in~\cite{sahin2019inexact,boumal2020deterministic}, and 
corresponds to the well-known
linear independence constraint qualifications (LICQs) for the feasibility problem
\begin{equation}
\textup{find }U \textup{ such that }\A(UU^\top)=b,
\label{eq:feas2}
\end{equation}
which attempts to find \underline{a} matrix $U$ with zero~\eqref{eq:fgDefn}, i.e., a matrix $U$ that satisfies $G(U)=0$. 
%  the feasibility problem~$\min_U G(U)$, which attempts to find a  matrix that satisfies the constraints of~\eqref{eq:bm}. 
%  More specifically, 
 Assumption~\ref{assumption:key}\ref{assumption:man} corresponds to the weakest sufficient conditions under which the KKT  conditions are necessary for feasibility
% global optimality 
in~\eqref{eq:feas2}, see~\cite{ruszczynski2011nonlinear,nocedal2006numerical}.
% the feasibility problem~$\min_U G(U)$.
% see~\eqref{eq:measOp} and~\eqref{eq:fgDefn} to review the notation. 
% Recall the definition of the feasibility gap~$G$ in~\eqref{eq:fgDefn}.
Without Assumption~\ref{assumption:key}\ref{assumption:man},  
% That is, Assumption~\ref{assumption:key} comprises the weakest sufficient conditions under which there can 
% in general there cannot be be any hope of efficiently  finding a matrix~$U$ that  achieves zero training error ($\A(UU^\top)=b$). That is, without Assumption~\ref{assumption:key}\ref{assumption:man}, 
the~\eqref{eq:fgDefn} is \underline{not} necessarily dominated by its gradient $(\nabla G(U)=0 \not\Rightarrow G(U)=0)$, in which case we cannot in general hope to solve \eqref{eq:feas2} in polynomial time with a  first-order optimization algorithm.

That is, without Assumption~\ref{assumption:key}\ref{assumption:man}, we cannot in general hope to efficiently find  \underline{any}  matrix that achieves zero training error! Such peculiarities are not uncommon in nonconvex optimization~\cite{murty1985some}.
From this perspective, Assumption~\ref{assumption:key}\ref{assumption:man} is \underline{minimal} in order to successfully recover the planted model $X^\n$. 

% Similar to \cite[Section~6]{boumal2020deterministic} and~\cite{eftekhari2020implicit}, Assumption~\ref{assumption:key} corresponds to the standard constraint qualifications for \eqref{eq:bm}, % see~\eqref{eq:fgDefn}, 
% which are the weakest sufficient conditions under which there can in general be any hope of tractably finding a matrix $U$ that agrees with the available data in the sense that $\A(UU^\top) = b$, see\eqref{eq:bm}.  Let us add that Assumption~\ref{assumption:key} is verified for a few examples in \cite[Section~4]{boumal2016non}, see also~\cite{sahin2019inexact} and the notion of geometric regularity therein.  

% later enables us to solve~\eqref{eq:feas} to global optimality and find \emph{a} rank-$p$ model with zero training error.
% see  
% Remark~\ref{rem:licqNeeded}. 
% later explains why  Assumption~\ref{assumption:key}\ref{assumption:man} is necessary in order to recover the planted model~$X^\n$. 

See also~\cite{boumal2016non,sahin2019inexact,boumal2020deterministic} for precedents of Assumption~\ref{assumption:key}\ref{assumption:man}. Beyond LICQs, the supplementary  relates 
Assumption~\ref{assumption:key}\ref{assumption:man} to other notions in optimization theory, e.g., the Polyak-{\L}ojasiewicz condition~\cite{karimi2016linear,polyak1963gradient}.  As a side note, if
Assumption~\ref{assumption:key}\ref{assumption:man} is fulfilled, then $\Int(\M_b)$ is a closed embedded submanifold of $\R^{d\times p}$ with co-dimension~$m$~\cite[Corollary~5.24]{lee2013smooth}. 
% which justifies its name. 
Here, $\Int(\cdot)$ stands for the relative interior of a set.

The appendix also establishes that Assumption~\ref{assumption:key}\ref{assumption:man} holds almost surely for the generic linear operator $\A$ that was described in Theorem~\ref{corr:mainSimple}, provided that $m=\Omega(\rank(U_0)d\log d)$, i.e., provided that the problem is sufficiently over-parametrized.

Beyond the common choice for the operator $\A$ in Theorem~\ref{corr:mainSimple},  verifying Assumption~\ref{assumption:key}\ref{assumption:man} is often difficult in practice, even though this assumption has several precedents within the nonconvex optimization literature, e.g.,~\cite{boumal2016non,sahin2019inexact,boumal2020deterministic}.  In this sense, Theorem~\ref{thm:main} should be  regarded primarily as a theoretical result that sheds  light, for the first time, on the role of initialization rank in the implicit bias of gradient flow in matrix sensing.   
\end{remark}

\begin{remark}[\textsc{Initialization rank}]\label{rem:initRank}
The RIP  in Assumption~\ref{assumption:key}\ref{assumption:iso}  is common  in matrix sensing and ensures that the operator $\A$ acts as an injective map, when its domain is restricted to  sufficiently low-rank matrices~\cite{davenport2016overview}. For example, the generic linear operator $\A:\R^{d\times d}\rightarrow\R^m$   in Theorem~\ref{corr:mainSimple} satisfies the $r$-RIP, provided that $m={\Omega}(rd\log d)$~\cite[Theorem~2]{geyer2020low}\cite{foucart2019iterative}.

% , i.e., provided that $m \gtrsim rd\log d$, where we ignored the constants~\cite[Theorem~2]{geyer2020low}\cite{foucart2019iterative}.

Considering our (often) limited sampling budget, we must in practice  select a sufficiently low-rank initialization for the~\eqref{eq:flowG} to ensure that Assumption~\ref{assumption:key}\ref{assumption:iso} is fulfilled. For example,  for the generic operator~$\A$  in Theorem~\ref{corr:mainSimple}, we must initialize the~\eqref{eq:flowG} at $U_0$ such that $\rank(U_0) =O(m/(d\log d))$.

Therefore, Assumption~\ref{assumption:key}\ref{assumption:man} correctly reflects the limitations of implicit bias of the~\eqref{eq:flowG}. Indeed, we observed numerically in Example~\ref{ex:numEx}  that,  with a high-rank initialization,  the~\eqref{eq:flowG}  is in general \underline{not} implicitly biased towards $X^\n$. The exception is  when the initialization is also nearly zero as in  \eqref{eq:nearlyZero}. 
% This latter near zero initalization is the focus of much of the literature of implicit bias in matrix sensing, see \eqref{eq:nearlyZero}.
% ~\cite{gunasekar2017implicit,arora2019implicit,li2018algorithmic}.

% (unless when the initialization norm is very small, which is studied in~\cite{li2018algorithmic}. 
% Our  discussion here also explains the difference between Assumption~\ref{assumption:key}\ref{assumption:iso} and its analogue in~\cite{li2018algorithmic}. The former assumes the $(2\rank(U_0))$-RIP, whereas the latter  assumes that $\A$ satisfies the $(4\rank(X^\n))$-RIP in order  to study the implicit bias of gradient flow in matrix sensing in the limit of \underline{vanishing initialization norm}. 
\end{remark}

%%%%%%%%%

% Let us discuss the assumptions of Theorem~\ref{thm:main}. We have already clarified in Section~\ref{sec:setup} that Assumption~\ref{assumption:key} is minimal in order to answer Question~\ref{q:q}. Item~\ref{item:nearbyCorr} in Theorem~\ref{thm:main} can be easily met by choosing $U_0$ sufficiently close to the origin (or alternatively by setting $\xi$ in~\eqref{eq:model} sufficiently large). Likewise, Item~\ref{item:initRankLarge} above can be easily fulfilled. 
% Below, we clarify Item~\ref{item:howSmallInitV2Corr}. 
%   We will also explain below how the informal statement of our result in the Introduction follows from Theorem~\ref{thm:main}.
   }

\begin{remark}[\textsc{Initialization norm}]\label{rem:initNorm}
An initialization $U_0$ that satisfies Assumptions~\ref{assumption:key}\ref{item:initRankLarge}-\ref{item:howSmallInitV2Corr} exists under mild conditions, e.g., by appealing to the Pataki's lemma~\cite{pataki1998rank,polik2007survey}. Note that Theorem~\ref{thm:main} is an example of a ``capture theorem'' in  nonconvex optimization~\cite{boumal2016non,sahin2019inexact,li2018calculus,attouch2013convergence}, i.e., Theorem~\ref{thm:main} predicates on an initialization  within a specific ``capture neighborhood'' of
the set $\M_b$  in~\eqref{eq:manifold}. Equivalently, Theorem~\ref{thm:main} predicates on an initialization within a sufficiently small neighborhood of the set of all matrices with zero training error, see~\eqref{eq:howSmallInitV2Corr}.
% the feasibility problem: find $U$ such that $G(U)=0$. 
As stipulated 
% by Assumption~\ref{assumption:key}\ref{item:howSmallInitV2Corr}, 
in~\eqref{eq:a}-\eqref{eq:b}, this  neighborhood also coincides with the set of all matrices with a sufficiently small training (or test) error.

Phrased differently,  Theorem~\ref{thm:main} applies only when the~\eqref{eq:flowG} is initialized sufficiently close to the set~$\M_b$ in~\eqref{eq:manifold}, see \eqref{eq:howSmallInitV2Corr}.
% the set of all matrices with zero training error $(G(U)=0)$. 
Equivalently, in light of~\eqref{eq:a} and \eqref{eq:b}, Theorem~\ref{thm:main} applies only when the  training or test error is sufficiently small at the initialization.

% Theorem~\ref{thm:main} is a ``capture theorem'', common in the literature of nonconvex optimization~\cite{sahin2019inexact,boumal2016non}, which predicates on an initialization within a specific ``capture neighborhood'' of the feasibility problem~\eqref{eq:licqRem}, as specified by~\emph{\ref{item:nearbyCorr}} and~\emph{\ref{item:howSmallInitV2Corr}}. To help form a better mental image, consider~\eqref{eq:howSmallInitV2Corr}, which clearly specifies a neighborhood of~$\M_b$, the set with zero training error.   Put differently, our results apply only when the~\eqref{eq:flowG} is initialized within this capture neighborhood. 

% Such capture theorems are fundamentally different from  local refinement results, appearing within the signal processing literature~\cite[Chapter 5]{chi2019nonconvex}. Note that the local refinement results rely on an initialization within a  small neighborhood of an isolated global minimizer where the objective function is locally strongly convex. 

Assumption~\ref{assumption:key}{\ref{item:howSmallInitV2Corr}} partially reflects  the limitations of implicit bias, and loosely agrees with our  numerical observation in Example~\ref{ex:numEx} that there is no implicit bias when the~\eqref{eq:flowG} is initialized far from the origin, see Figure~\ref{fig:break}. 
% Indeed, 
% % n the following sense. 
% if Assumption~\ref{assumption:key}\ref{item:howSmallInitV2Corr} is violated  and the initialization norm is large $(\|U_0\|\gg 0)$, then it is well-known that implicit regularization towards $X^\n$ might disappear, see Figure~\ref{fig:break} for a numerical example. 
To illustrate this connection, let us focus only on~\eqref{eq:a}.
Note that 
\begin{align}
     \|U_0\|^2 = \|U_0U_0^\top\| & \ge \|U_0U_0^\top -X^\n\| - \|X^\n\| \ge \|U_0U_0^\top-X^\n\|-\xi \nonumber  \\ 
     & \ge \vertiii{\A}^{-1}\cdot \|\A(U_0U_0^\top-X^\n)\|_2-\xi  =\vertiii{\A}^{-1} \cdot \|\A(U_0U_0^\top) - b\|_2 -\xi, 
    \label{eq:reverseTriIneq}
\end{align}    
where~$\vertiii{\A}$ is the appropriate operator norm of $\A$, and we repeatedly used~\eqref{eq:model} and~\eqref{eq:measOp} above. In view of~\eqref{eq:reverseTriIneq}, if the initial training error is large $(\|\A(U_0U_0^\top) - b\|_2 \gg \vertiii{\A} \xi)$, then the initialization norm will be large $(\|U_0\|_\F\gg 0)$ and the~\eqref{eq:flowG} will in general show no implicit bias towards the planted model $X^\n$, as we saw numerically in Figure~\ref{fig:break}. 
We lastly note that Assumption~\ref{assumption:key}\ref{item:howSmallInitV2Corr} is difficult to verify in practice
% , as  discussed further in Remark~\ref{rem:practicalInit}, 
and, as discussed in Remark \ref{rem:manifold},  Theorem~\ref{thm:main} should therefore be regarded primarily as a theoretical result that sheds light on the role of initialization rank in implicit bias.  
% rather than a practical initialization technique for the~\eqref{eq:flowG}. 
% To complement this theoretical result, we will later suggest an adaptive initialization  heuristic for the~\eqref{eq:flowG}, which might be of independent interest.
\end{remark}

\begin{remark}[\textsc{Local refinement}]
Capture theorems, such as Theorem~\ref{thm:main}, are fundamentally different from the local refinement results within the signal processing literature. Such local refinement results require us to initialize the \eqref{eq:flowG} within a very small neighborhood of a matrix with zero test error, in which local strong convexity holds~\cite[Chapter 5]{chi2019nonconvex}. In contrast, our capture neighborhood in~\eqref{eq:a} contains \underline{all} matrices with zero training error (which also of course includes all matrices with zero test error). In fact, we observe from \eqref{eq:a} that the capture neighborhood in Theorem~\ref{thm:main} is a ``tube'' of the form $\{U:\|\A(UU^\top-\underline{U}\,  \underline{U}^\top) \|_2\le \rho\}$ for a certain radius $\rho$,  rather than a small Euclidean ball centered at a matrix  $\underline{U}$ that has zero test error $(\underline{U}\, \underline{U}^\top = X^\n)$.  
% Here, $a$ is a geometric attribute of the problem, which will be made precise later. 
\end{remark}

\begin{remark}[\textsc{Proof of Theorem} \ref{corr:mainSimple}] To establish Theorem \ref{corr:mainSimple} as a corollary of Theorem \ref{thm:main}, it suffices to verify that Assumption \ref{assumption:key} is fulfilled. Because $p=d$ in Theorem \ref{corr:mainSimple}, Assumption \ref{assumption:key}\ref{assumption:infoLevel} is met trivially, possibly after infinitesimally perturbing $X^\n$, see Remark \ref{rem:recoverable}. Assumptions \ref{assumption:key}\ref{assumption:man}-\ref{assumption:iso} hold with overwhelming probability in view of Theorem \ref{corr:mainSimple}\ref{item:iso} and Remarks \ref{rem:manifold} and \ref{rem:initRank}. Assumption \ref{assumption:key}\ref{item:initRankLarge} holds by Theorem \ref{corr:mainSimple}\ref{item:initRank} and $\effrank\le \rank$. Lastly, Assumption~\ref{assumption:key}\ref{item:howSmallInitV2Corr} holds by Theorem \ref{corr:mainSimple}\ref{item:nearby}-\ref{item:nearby2}. 

\end{remark}

\vspace{-5pt}
\section{Related Work}\label{sec:relatedWork}
\vspace{-5pt}

%  {In this section, we review the available answers to Question~\ref{q:q}.}
%  {When $p=d$ and the~\eqref{eq:flowG} is initialized near the origin ($U_0\approx 0$),~\cite{li2018algorithmic} showed that $X^\n$ can be recovered,
% %  Question~\ref{q:q} was answered affirmatively in~\cite{li2018algorithmic}, 
%  if the operator~$\A$ satisfies the RIP. 
 Under various assumptions on the operator~$\A$ in \eqref{eq:measOp},~\cite{gunasekar2017implicit,arora2019implicit,li2018algorithmic} have  studied the implicit bias of the~\eqref{eq:flowG} in matrix sensing, see also \cite{ma2018implicit,tu2016low,woodworth2020kernel,gidel2019implicit,geyer2020low}.
In particular, for a specially designed operator~$\A$, for which~$\M_b$ in~\eqref{eq:manifold} aligns with the level sets of the convex matrix sensing problem,~\cite{geyer2020low} proved that the planted model~$X^\n$ is the unique PSD matrix that satisfies $\A(X^\n)=b$. 
% even though $m=\wt{\Omega}(rd)$. 
% as opposed to matrix sensing. 
For our purposes, the most relevant past result is Theorem~1.1 in~\cite{li2018algorithmic}, 
% simplified below and 
adapted below to our setup.
% with gradient flow.
\begin{thm}[\textsc{State of the art, simplified}]\label{thm:srebro} 
Consider the framework specified by~\eqref{eq:model} and~\eqref{eq:measOp}. 
Suppose that the operator $\A$ satisfies the $(4\cdot \rank(X^\n))$-RIP. Suppose also that~$p=d$, and that the~\eqref{eq:flowG} is initialized at $U_0 =u_0 I_d\in\R^{d\times d}$ for a sufficiently small~$u_0>0$, where $I_d$ denotes the identity matrix.  Then it holds that 
\begin{align}
    \|U(t)U(t)^\top - X^\n\|_\F^2 \lesssim u_0 \sqrt{d}/ \kappa(X^\n)^2,
    \qquad \forall t\in [\kappa(X^\n) \log(d/u_0), 1/\sqrt{u_0 d\cdot \kappa(X^\n) }]
\end{align}
where $\kappa(X^\n)$ is the condition number of $X^\n$, i.e., the ratio of its largest and smallest nonzero singular values. Above, $\lesssim$ suppresses any unnecessary factors. 
\end{thm}

Note that Theorem~\ref{thm:srebro} is limited to the case $p=d$, and 
% does {not}
% affirmatively answer Question~\ref{q:q},
% in view of its lack of convergence. Indeed,  Theorem~\ref{thm:srebro} does 
% because it 
does \emph{not} guarantee the convergence of the~\eqref{eq:flowG}
% $U(t)U(t)^\top$ 
to the planted model~$X^\n$.
Intuitively, when initialized near the origin, the~\eqref{eq:flowG} moves rapidly along the ``signal'' directions and $U(t)U(t)^\top$  approaches a small neighborhood of~$X^\n$. After this initial phase, the contribution of ``noise'' directions might potentially accumulate and push~$U(t)U(t)^\top$ away from the planted model~$X^\n$. 
% it is worth noting that 
(We  never observed this divergence numerically.)

There are more fundamental differences between our Theorem \ref{thm:main} and Theorem~\ref{thm:srebro} from~\cite{li2018algorithmic}. In the latter, the initialization is full-rank and near-zero. In contrast, as discussed in Remarks \ref{rem:initRank} and~\ref{rem:initNorm}, our Theorem \ref{thm:main} applies to a low-rank initialization with a sufficiently small training (or test) error. In Remarks~\ref{rem:initRank} and~\ref{rem:initNorm}, we also explained that our requirements on the initialization rank and norm (partially) mirror the true {limitations} of implicit bias that we had earlier identified in Example \ref{ex:numEx}. 

Theorem \ref{thm:main} can also be used to slightly improve upon Theorem \ref{thm:srebro}: For a sufficiently small  $u_0$  and a sufficiently large $t_0$, suppose that $\rank(U(t_0))\le 2\rank(X^\n)$. Then it is easy to see that Theorem \ref{thm:srebro} can be strengthened to $\lim_{t\rightarrow\infty}\|U(t)U(t)^\top- X^\n\|_\F=0$, provided that Assumption~\ref{assumption:key}\ref{assumption:man} holds.

This work identifies the limitations of the implicit bias of gradient flow in matrix sensing. For limitations of implicit bias in other settings, among others, see \cite{vardi2020implicit} for  shallow neural networks, \cite{dauber2020can} for stochastic convex optimization, and~\cite{razin2020implicit}
for matrix completion, where the RIP never holds.

\vspace{-5pt}
 \section{A New Algorithm for Matrix Sensing}\label{sec:newAlg}
 \vspace{-5pt}
 
We  emphasize that Theorem~\ref{thm:main} primarily provides theoretical insights, 
    % and it \underline{does} provide a  practical initialization scheme for the~\eqref{eq:flowG}. Instead, Theorem~\ref{thm:main} 
% shedding  light on the  
% theoretical aspects of implicit bias of the~\eqref{eq:flowG} and, specifically its 
shedding light for the first time on the role of the initialization rank in
 implicit bias.
    % Indeed, it is in general not trivial to efficiently find the initialization $U_0$ prescribed in Assumption~\ref{assumption:key}\ref{item:howSmallInitV2Corr}.
    To complement Theorem~\ref{thm:main}, we next propose an adaptive heuristic for matrix sensing that replaces Assumption~\ref{assumption:key}\ref{item:howSmallInitV2Corr} with a more practical guideline. This new algorithm complements the full-rank and near-zero initialization scheme that has dominated  the existing literature of implicit bias in matrix sensing~\cite{gunasekar2017implicit,arora2019implicit,li2018algorithmic}, see \eqref{eq:nearlyZero}. 
    Beyond matrix sensing, a near-zero initialization often leads to vanishing gradients in deep learning~\cite{hochreiter2001gradient}. In that sense, alternatives to \eqref{eq:nearlyZero}, such as the  new algorithm below, might in the future provide valuable insights for more difficult learning problems.   

\begin{figure}%[h!]
        \centering
        \includegraphics[width=6.9cm,height=4.5cm]{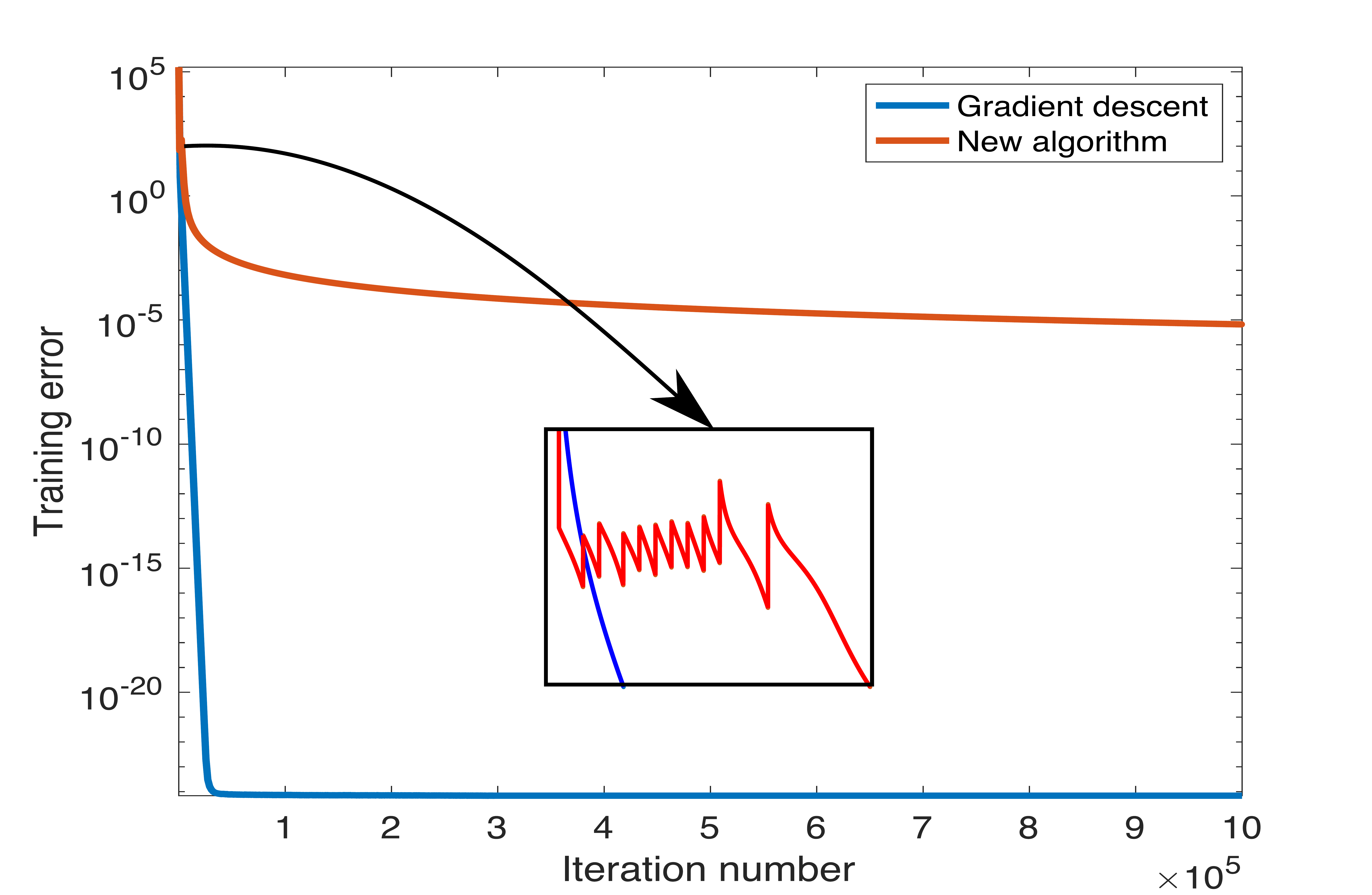} 
        \hfill
        \includegraphics[width=6.9cm,height=4.5cm]{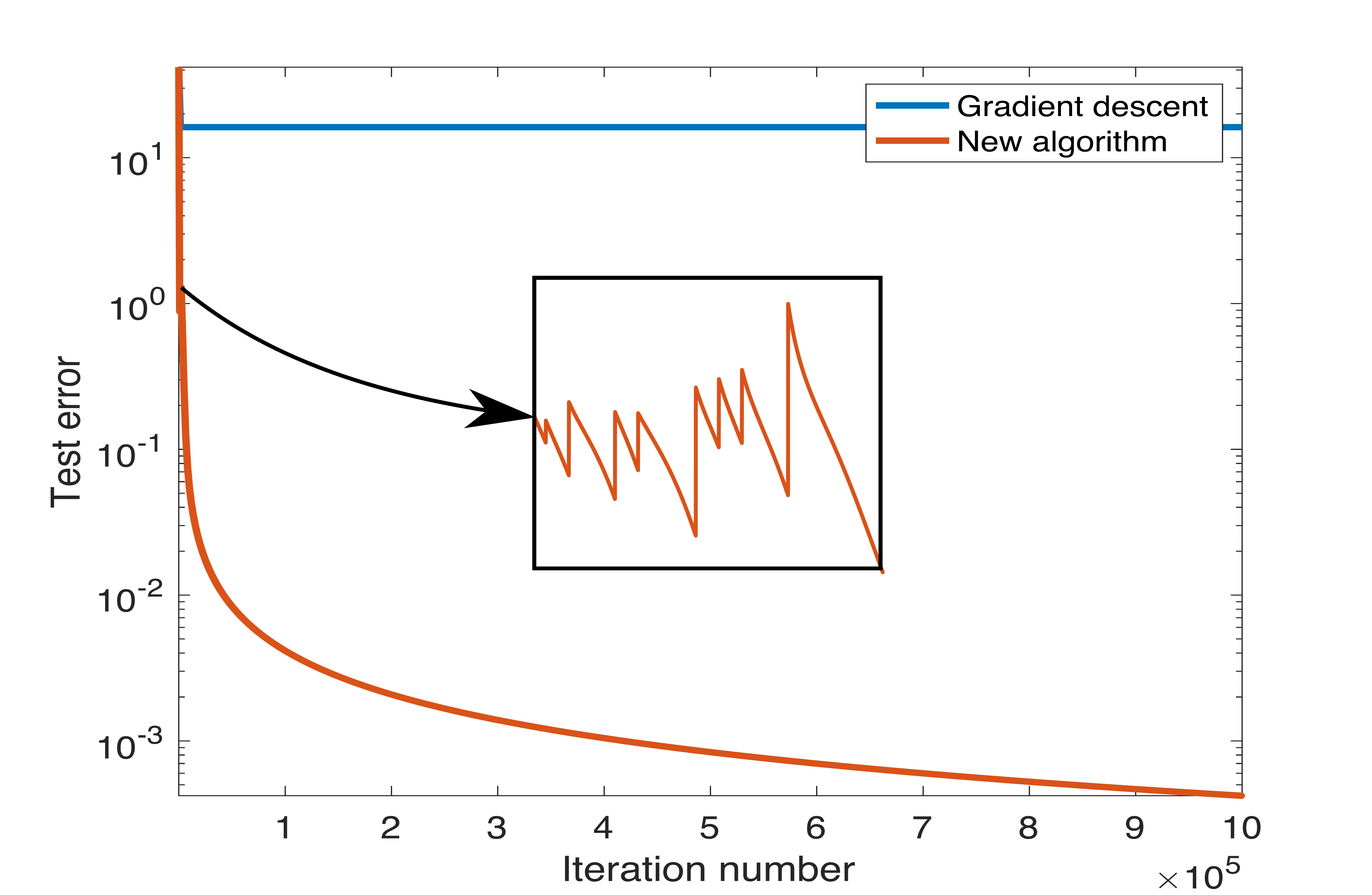}
        \caption{
  The restarts of the new algorithm are magnified, see Section \ref{sec:newAlg} and the MATLAB code. 
        }\label{fig:test} 
    \end{figure}

    The message of Example \ref{ex:numEx} and Theorem~\ref{thm:main} is that high-rank and high-norm initializations are in general both detrimental to implicit bias. 
    Another remarkable pattern which emerges from an inspection of  Figures~\ref{fig:lowNorm} and~\ref{fig:hiNorm} is that the test error is always large (no implicit bias), whenever the training error reduces too rapidly.
    % uuexponentially fast. 
% As a future research direction, 
The previous two observations naturally suggest  an adaptive heuristic for matrix sensing which restarts the training whenever the convergence is too fast. 
% This new algorithm is detailed in the supplementary material. 
 Figure~\ref{fig:test} illustrates a typical outcome, where  the new algorithm considerably outperforms the gradient descent~\eqref{eq:gradDec} with the same initialization.
The setup is identical to Example~\ref{ex:numEx}, but the random data is fresh and $\|U_0\|_\F=10$, $\eta=5\cdot 10^{-6}$, $W=100$, $\tau=0.998$, $r_0=30$, $\D_{\rank}=3$, $\text{factor}=1/2$, $r=2$.

% Restart the training if the training error decays too rapidly. 

% The gist of our work is that high-rank and faraway initialization is bad. ... 

% A comprehensive numerical investigation is left as future work but note that this adaptive initialization technique might also be of interest beyond matrix sensing. 

\noindent
   \begin{minipage}{1\linewidth}
\begin{algorithm}[H]

\textbf{Input:} 
The framework specified by \eqref{eq:model} and \eqref{eq:measOp},  $\eta>0$, $K,W\in \mathbb{N}$, $W\le K $, $\tau\ge0$,  $r_0\le d$, $\rho_0\ge 0$, $\Delta_{\text{rank}}\in \mathbb{N}$, $\text{factor}\in (0,1)$, $r\ge \rank(X^\n)$.
% step size $\eta$, scope $K\in \mathbb{N}$, window length $W\le K $, threshold~$\tau\ge0$, initialization rank $r_0\le d$ and norm $\rho_0\ge 0$, $\Delta_{\text{rank}}\in \mathbb{N}$ and $\text{factor}\in (0,1)$.

% Symmetric $d\times d$ matrices $\{A_i\}_{i=1}^m$ and the corresponding operator $\A$ in~\eqref{eq:measOp}, vector $b\in \R^m$, integer $p$ such that $pd\ge m$, initialization $U_0\in \R^{d\times p}$, positive penalty weight $\g$  and step sizes $\{\eta_k\}_k$. 

\vspace{0pt}

\noindent {Set $U_0\in \R^{d\times p}$ to be a rank-$r_0$ matrix inside $\ball(\rho_0):=\{U:\| U\|_{\F}\le \rho_0\}$.  

For $k=0,1,\cdots,K$, repeat
\begin{enumerate}[wide]

\item $U_{k+1} = U_k - \eta \nabla \|\A(U_kU_k^\top) - b\|_2^2$, \Comment{Gradient descent update}

\item  If $\mod(k,W)= 1$, then \Comment{Every $W$ iterations, }
\begin{enumerate}
    \item  $\text{ratio} \leftarrow \underset{k-W+1 \le k'\le k}{\max}  \frac{\|\A(U_{k'}U_{k'}^\top)-b\|_2^2}{\|\A(U_{k'-1}U_{k'-1}^\top)-b\|_2^2}$
    \item If $\text{ratio} < \tau$  \Comment{if convergence was linear in the last window, then}
    \begin{itemize}[wide]
        \item  $\rho_0 \leftarrow  \text{factor} \cdot \rho_0$ and $r_0\leftarrow \max(r_0 - \Delta_{\rank},r) $ \Comment{reduce the initialization norm and rank,}
        \item Set $U_{k+1}\in \R^{d\times p}$ to be a rank-$r_0$ matrix inside $\ball(\rho_0)$. \Comment{and restart the training.}
    \end{itemize}
\end{enumerate}

\end{enumerate}
}

\caption{
A new algorithm for matrix sensing
\label{fig:alg}}
\end{algorithm}
       
\end{minipage}
    \hfill

\bibliographystyle{unsrt}
\bibliography{References}

\appendix

\section{Geometry of the Set $\M_b$}
\label{sec:nonConvexGeom}

% To begin, this section studies the set of matrices that agree with the training data. More specifically, 

This section studies \editae{with more depth} the  geometry of the set $\M_b$   {and its neighborhood in $\R^{d\times p}$}. \editae{Let us recall that~$\M_b$ is the set of all matrices with zero training error and bounded norm, introduced  in~\eqref{eq:manifold} and briefly discussed in Section~\ref{sec:setup}}.
% For brevity, we will often denote the compact feasible set of problem~\eqref{eq:bm} by 
% \begin{align}
%     {\M_b} & := \{ U: \A(U U^\top) = b,\, \|U\|\le  \xi \} \nonumber\\
%     & = \{U: g(U) = 0,\, \|U\|\le   \xi \} \subset \R^{d\times p}, \,\, \text{(see \eqref{eq:fgDefn})}
%     \label{eq:manifold}
% \end{align}
% so that problem~\eqref{eq:bm} is also equivalent to $\min_U\, f(U) $ subject to $U\in \M_b$. 
Recalling  {the definition of $g$ in}~\eqref{eq:fgDefn}, note  {also} that the {(total)} derivative of $g(U)$ at $U$ is the linear operator  $\Der g(U):\R^{d\times p} \rightarrow\R^m$, defined as
\begin{align}
    \Der g(U)[\D] := \A(\D U^\top),
    \label{eq:defnC}
\end{align}
and its adjoint  {operator} is specified as
\begin{equation}
    (\Der g(U))^* [\d] := \A^*(\d)\cdot U = \sum_{i=1}^m \d_i A_i U,
        \label{eq:adjO}
\end{equation}
where $\A^*$ is the adjoint of the linear operator $\A$ in~\eqref{eq:measOp} and $\d_i$ is the $i^{\text{th}}$ entry of the vector~$\d\in \R^m$. 
% The following assumption is central to our analysis,  similar to~\cite[Assumption~1.1]{boumal2020deterministic}.  
% Remark~\ref{rem:licqNeeded}  later explains why  {the} assumption  {below} is necessary in order to
% achieve Objective~\ref{o:objective}  {and affirmatively answer Question~\ref{q:q}.}   {Note also that the assumption  below subsumes the earlier Assumption~\ref{assumption:operator}.} 
% \begin{assumption}[Manifold]\label{assumption:key}Assumption~\ref{assumption:operator} is fulfilled and there also exists an integer $m'\le \rank(X^\n) d$ such that
% \begin{align}
%     \rank(\Der g(U)) = m', 
%     \label{eq:fullRankAll}
% \end{align}
% for every $U$ in an open neighborhood of $\M_b$ in~\eqref{eq:manifold}.
% Equivalently,  {the matrices}~$\{A_i U\}_{i=1}^m$ span an $m'$-dimensional subspace of $\R^{d\times p}$, for every $U$ in an open neighborhood of~$\M_b$. 
% \end{assumption}
Under Assumption~\ref{assumption:key}\ref{assumption:man},  $\Int(\M_b)$ is a closed embedded submanifold of $\R^{d\times p}$ of co-dimension $m$, where $\Int(\cdot)$ stands for the relative interior. 
Beyond LICQs in Remark \ref{rem:manifold}, Assumption~\ref{assumption:key}\ref{assumption:man} is also related to the Polyak-{\L}ojasiewicz (PL) condition~\cite{karimi2016linear,polyak1963gradient}.  If the compact set $\M_b$ in~\eqref{eq:manifold} satisfies Assumption~\ref{assumption:key}\ref{assumption:man}, there exists~$\mu>0$ such that~$G$ in~\eqref{eq:fgDefn} satisfies 
\begin{align}
\frac{1}{2}\| \nabla G(U)\|_\F^2 & = \frac{1}{2} \l\| (\Der g(U))^*[g(U)] \r\|_\F^2 
\qquad \text{(chain rule)}
\nonumber\\
& \ge \frac{\mu}{2} \|g(U)\|_2^2 = \mu \cdot G(U),
% \quad \text{(see \eqref{eq:fgDefn})}
\label{eq:pl}
\end{align}
for every $U$  {in an open neighborhood of~$\M_b$.} In view of~\eqref{eq:pl},   Assumption~\ref{assumption:key}\ref{assumption:man} implies the~PL condition for the function $G$, when restricted to the  {a} neighborhood of $\M_b$, see~\cite[Equation (3)]{karimi2016linear}.   In passing, we  {also} remark that   {Assumption~\ref{assumption:key}} also  {relates} to the Mangasarian-Fromovitz and Kurdyka-{\L}ojasiewicz conditions~\cite{bolte2018nonconvex,xu2017globally,sahin2019inexact}. Next, consider  the generic operator $\A$ described in Theorem \ref{corr:mainSimple}. We assume without loss of generality that~$m\le d^2$ because the problem is otherwise trivial.
Consequently, for fixed $U\in \R^{d\times d}$, $\{A_iU\}_{i=1}^m \subset \R^{d\times d}$ are almost surely linearly independent, thanks to the random design of $\A$. Since $\M_b$ in \eqref{eq:manifold} is a compact set, it also follows that Assumption \ref{assumption:key}\ref{assumption:man} is fulfilled almost surely.

 {Let us next study}  the neighborhood of the  set 
$\M_b$ in~\eqref{eq:manifold}, beginning  {below} with a   {more} quantitative approach to
 Assumption~\ref{assumption:key}\ref{assumption:man}.  {More specifically, below we  define a well-behaved neighborhood of the set~$\M_b$ in~\eqref{eq:manifold}, the existence of which will shortly be guaranteed under Assumption~\ref{assumption:key}\ref{assumption:man}. We add that this specific neighborhood of~$\M_b$, defined below, is of key importance for us. Indeed, as we will  see later in this section, any limit point of the~\eqref{eq:flowG}, within this well-behaved neighborhood, achieves zero training error.}

\begin{defn}[Geometric regularity]\label{defn:GR}
{Suppose that Assumption~\ref{assumption:key}\ref{assumption:man} holds and fix $\rho\in [0,\infty)$.}  {Recall the map~$g$ in~\eqref{eq:fgDefn}.}
 {We say that the} set $\M_b$ in~\eqref{eq:manifold} satisfies the $(\rho,m)$-geometric regularity or $(\rho,m)$-GR if 
\begin{equation}
% \nu_\rho & = 
\min \l\{ \s_{m}(\Der g (U)) : \dist(U,{\M_b}) \le \rho
 \r\} 
  >0, 
 \label{eq:defnReg}
\end{equation}
where $\s_{i}(\cdot)$  returns the $i^{\text{th}}$ largest singular value of a linear operator, and
\begin{equation}
\dist(U,{\M_b}) := \min_{U'\in {\M_b}}\|U-U'\|_{\F}
\label{eq:metricDefn}
\end{equation}
is the distance from  {the matrix} $U\in \R^{d\times p}$ to the compact set ${\M_b}$.
\end{defn}

 {Before we proceed, for completeness, a short remark follows next to justify the choice of metric in Definition~\ref{defn:GR}.
\begin{remark}[Invariance of the metric] 
The metric in~ {\eqref{eq:metricDefn}} is  invariant under rotation from right. That is,  for any $U\in \R^{d\times p}$ and $R\in \O_p$, it holds that 
\begin{align}
    \dist(U R ,\M_b) & = \min_{U'\in \M_b} \|U R - U'\|_\F \nonumber\\
    & = \min_{U'\in \M_b} \|U R - U'R \|_\F \nonumber\\
    & = \min_{U'\in \M_b} \|U - U'\|_\F \nonumber\\
    & = \dist(U ,\M_b),
\end{align}
where the second line above holds because  $U'\in \M_b$ implies that $U'R\in \M_b$, see~\eqref{eq:manifold}. The third line  holds by the rotational invariance of the Frobenius norm.  Here, $\O_p =\{R: R^\top R = I_p\}\subset \R^{p\times p}$ denotes the orthogonal group and~$I_p\in \R^{p\times p}$ is the identity matrix. 

\end{remark}}

Using a  {standard}  perturbation argument, the next result establishes that the set $\M_b$ satisfies the  {geometric regularity}, 
 {provided that Assumption~\ref{assumption:key}\ref{assumption:man} is fulfilled}. That is, simply put, Definition~\ref{defn:GR} is not vacuous.  
%  {That is, the result below calculates (a lower bound for) the radius of the well-behaved neighborhood of the set $\M_b$ specified in Definition~\ref{defn:GR}.}

\begin{prop}[Geometric regularity] \label{prop:GRHolds} If Assumption~\ref{assumption:key}\ref{assumption:man}  {is met}, then the set $\M_b$ in~\eqref{eq:manifold} satisfies the $(\rho_0,m)$-GR with 
\begin{align}
  \rho_0:=  \frac{\s_{m}({\M_b})}{2\|\A\|},  
% \qquad \s_{m'}({\M_b}):= \min_{U\in {{\M_b}}}\,\, \s_{m'}(\Der g(U)) > 0.
\label{eq:sMinDefn}
\end{align}
where $\s_{m}({\M_b}):= \min\{\s_{m}(\Der g(U)): U\in {{\M_b}}\}  > 0$ and $\|\A\|$ is the operator norm of $\A$ in~\eqref{eq:measOp}.  {Here, $\s_{m}(\Der g(U))$ returns the $m^{\text{th}}$ largest singular value of the operator $\Der g(U)$.}
\end{prop}

Let us  record below  a consequence of~\eqref{eq:pl} and Proposition~\ref{prop:GRHolds},  {which will be later central to the proof of the main result of this paper.}  {In words, the  lemma below posits that the  {(nonconvex)} problem~\eqref{eq:feas} does not have any spurious first-order stationary points within a neighborhood of the set $\M_b$ in~\eqref{eq:manifold}. In particular, the lemma below implies that any limit point of the~\eqref{eq:flowG} within this neighborhood has zero training error. }

\begin{lem}[\textsc{No spurious FOSP}]\label{lem:noSpFOSP}  Suppose that Assumption~\ref{assumption:key}\ref{assumption:man} holds. Then any  {first-order stationary point (FOSP)} of  $G$ inside the set $\{U:\dist(U,\M_b)\le \rho_0\}$ is also a global minimizer of  {the problem~\eqref{eq:feas} (zero training error)}. That is, if $\ol{U}\in\R^{d\times p}$ satisfies  $\nabla G(\ol{U})=0$ and  $\dist(\ol{U},\M_b)\le \rho_0$, then $\A(\ol{U}\, \ol{U}^\top) = b$.
% or equivalently $\ol{U}\in \M_b$, 
See~\eqref{eq:measOp},~\eqref{eq:fgDefn}  and~\eqref{eq:sMinDefn}  {to review the notation used in this lemma}. 
\end{lem}

% \section{$\Int(\M_b)$ is a Manifold}

% Let $B_\xi := \{U: \|U\|_\F<\xi\} \subset \R^{d\times p}$ denote the (open) ball of radius  $\xi$ with respect to the Frobenius norm and centered at the origin.  Under Assumption~\ref{assumption:key}, $\Int(\M_b) = \M_b \cap B_\xi$ in~\eqref{eq:manifold} is a closed embedded submanifold of $\R^{d\times p}$ by~\cite[Theorem 5.22]{lee2013smooth}. Here, in the notation of the book, we set  $M=B_\xi $, $N=\R^m$, and $F$ is a smooth map such that $\Der F(U)$ is a  rank-$m$ linear operator for every $U\in M$ and $F$ agrees with $g$ in an open neighborhood of $\M_b$ with respect to the topology of $M$.

\section{Proof of Proposition~\ref{prop:GRHolds}}\label{sec:proofGRHolds} For $U\in \R^{d\times p}$, let $U_{\M_b}\in {\M_b}$ be the projection of $U$ onto ${\M_b}$, i.e., 
\begin{align}
\text{dist}(U,{\M_b}) = \|U - U_{\M_b}\|_{\F} \le \|U - U'\|_{\F},
\qquad \forall U'\in {\M_b}.
\label{eq:decomposition1}
\end{align} 
Above, note that $U_{\M_b}$ exists by the compactness of ${\M_b}$ in~\eqref{eq:manifold}, but might not be unique. 
% In view of the optimality of $u_{\M_b}$ above and the differentiability of the manifold ${\M_b}$ by Assumption~\ref{assumption:key}, there exists $\D\in \R^{d\times p}$ such that 
% \begin{align}
% u = u_{\M_b}+ \D,
% \qquad u_{\M_b} \in \Int(\M_b), 
% \qquad \D\in \N_{u_{\M_b}} {\M_b}, 
% \qquad 
% \| \D\|_F \ge \text{dist}(u,{\M_b}). 
% \label{eq:cndsDelta}
% \end{align}
% Note that $\D$ is the chord connecting $u$ to its projection $u_{\M_b}$ on  ${\M_b}$. 
Using the Weyl's inequality, note also that 
\begin{align}
    \s_{m}(\Der g(U)) & \ge \s_{m}(\Der g(U_{\M_b})) - \|\Der g(U) - \Der g(U_{\M_b})\|.
    % \qquad \text{(Weyl's inequality)}
    \label{eq:weylLayer}
\end{align}
To compute the operator norm of $\Der g(U) - \Der g(U_{\M_b})$ on the far-right above, we note that 
\begin{align}
    \| (\Der g(U) - \Der g(U_{\M_b}))[\D]\|_2 & = \|\A(\D U^\top) - \A(\D U_{\M_b}^\top)\|_2  \qquad \text{(see \eqref{eq:defnC})} \nonumber\\
    &  = \|\A ( \D (U - U_{\M_b})^\top )\|_2 
    \qquad (\A \text{ is linear})
    \nonumber\\
    & \le  \|\A\|\cdot  \|\D (U-U_{\M_b})^\top \|_\F  \nonumber\\
    & \le \|\A\| \cdot \|\D\|_\F \|U-U_{\M_b}\| \nonumber\\
    & \le \|\A\| \cdot \|\D\|_\F \|U-U_{\M_b}\|_\F \nonumber\\
    & = \|\A\| \cdot \|\D\|_\F \dist(U,\M_b), \qquad \text{(see \eqref{eq:decomposition1})}
    \label{eq:normDiffOperator}
\end{align}
for an arbitrary $\D \in \R^{d\times p}$. It follows from~\eqref{eq:normDiffOperator}  that 
\begin{align}
    \|\Der g(U) - \Der g(U_{\M_b}) \| \le \|\A\| \dist(U,\M_b). \label{eq:bndOnMaxDiff}
\end{align}
% Recalling \eqref{eq:adjO}, $(\Der g(u) - \Der g(u_{\M_b}))^*: \R^{d\times p}\rightarrow \R^m$ is defined as   
% \begin{align}
%     (\Der g(u) - \Der g(u_{\M_b}))^*[\d] = A^*(\d)\cdot (u-u_{\M_b}) = A^*(\d)\cdot \D.
%     \qquad \text{(see (\ref{eq:adjO},\ref{eq:cndsDelta}))}
% \end{align}
Using~\eqref{eq:bndOnMaxDiff}, we can  simplify~\eqref{eq:weylLayer} as 
\begin{align}
    \s_{m}(\Der g(U)) & \ge \s_{m}(\Der g(U_{\M_b})) - \|\A\| \dist(U,\M_b) 
    \qquad \text{(see (\ref{eq:bndOnMaxDiff}))}
    \nonumber\\
    & \ge  \s_{m}({{\M_b}}) - \|\A\| \dist(U,\M_b).
    \qquad \text{(see (\ref{eq:sMinDefn}))}
    \label{eq:nonsingularNeigh}
\end{align}
It immediately follows from \eqref{eq:nonsingularNeigh} that 
\begin{align}
      \min\l\{ \s_{m}(\Der g(U)) : 
    \dist(U,\M_{b}) \le \rho_0:= \frac{\s_{m}({{\M_b}})}{2\|\A\|}  \r\}
     \ge  \frac{\s_{m}({{\M_b}}) }{2} >0 , 
     \label{eq:defnNeigh}
\end{align}
% In words, the operator $\Der g(u)$ remains  nonsingular within the set $\M_{b,\rho}$. Note that $\M_{b,\rho}$  indeed contains a neighborhood of ${\M_b}$, i.e., 
% \begin{align}
%     \M_{b,\rho} \supseteq 
%     \l\{ u: \dist(u,{\M_b}) \le \rho \r\}
%     \qquad \text{(see \eqref{eq:cndsDelta})}
%     ,
% \end{align}
which completes the proof of Proposition~\ref{prop:GRHolds}.

\section{Proof of Theorem \ref{thm:main}}
% \label{sec:mainResults}

Here, we will  {use Lemma~\ref{lem:noSpFOSP} to} prove that  {the}~\eqref{eq:flowG}, if initialized properly, converges  to a limit point 
% $\ol{U}\in \R^{d\times p}$  such that $\ol{U}\cdot \ol{U}^\top $ is infinitesimally close to $X^\n$,
 {with zero test error.}
The first lemma in this section states that rank  does not increase along  {the}~\eqref{eq:flowG}.

\begin{lem}[Rank of gradient flow]\label{lem:rankInvariance} For an initialization $U_0\in \R^{d\times p}$,  {the}~\eqref{eq:flowG} satisfies 
\begin{align}
\rank(U(t))\le \rank(U_0), \qquad  t \ge 0.
\label{eq:rankInvEq}
\end{align}

\end{lem}

 {\emph{Proof sketch of Lemma~\ref{lem:rankInvariance}.}
The claim in Lemma~\ref{lem:rankInvariance} follows from writing the analytic singular value decomposition~(SVD) of $U(t)U(t)^\top$, then taking its derivative with respect to time $t$, and finally observing that any zero singular value of~$U(t)U(t)^\top$ remains zero throughout time. The detailed proof is deferred to the appendix.\hfill$\blacksquare$ 

The next  component of our argument posits that the~\eqref{eq:flowG} is bounded.} 
 {More specifically,} when initialized  {near} the set $\M_b$ in~\eqref{eq:manifold},  {the~\eqref{eq:flowG} always} remains near $\M_b$, as detailed next.

\begin{lem}[\textsc{Flow remains nearby}]\label{lem:flowNearby}
  Consider an initialization $U_0\in \R^{d\times p}$ such that  {$\|U_0\|\le \xi$ in~\eqref{eq:model}, that} $\rank(U_0)\ge \effrank(X^\n)$ and  {lastly that} $\dist(U_0,\M_b)$ is not too large,  {as specified precisely in~\eqref{eq:howSmallInit}.}
%   that $\xi$ in~\eqref{eq:model} 
%    {satisfies $\xi \ge \|U_0\|$}, and that 
Suppose  that Assumption~\ref{assumption:key}\ref{assumption:man} holds. Suppose also that  Assumption~\ref{assumption:key}\ref{assumption:iso} holds with $r\ge {2\cdot \rank(U_0)}$.   
   Then  {the}~\eqref{eq:flowG} satisfies 
   \begin{align}
    \dist(U(t),\M_b) < \rho_0, \qquad  t\ge 0,
    \label{eq:flowRemainsCloseAlways}
\end{align}
 {where $\rho_0$ was defined in~\eqref{eq:sMinDefn}.}
\end{lem}

 {\emph{Proof sketch of Lemma~\ref{lem:flowNearby}}.
The boundedness of the~\eqref{eq:flowG}, claimed in Lemma~\ref{lem:flowNearby}, follows from two observations: \circled{1}~$G(U(t))=\frac{1}{8}\|\A(U(t)U(t)^\top)-b\|_2^2$ in~\eqref{eq:fgDefn} is evidently bounded because the~\eqref{eq:flowG} moves along the descent direction~$-\nabla G(U(t))$. \circled{2}~Under Assumption~\ref{assumption:iso}, the restricted injectivity of the operator $\A$ allows us to translate the boundedness of~$G(U(t))$ into the boundedness of $U(t)$ in~\eqref{eq:flowG}.  A crucial ingredient of the proof is that Lemma~\ref{lem:rankInvariance} is in force  and, consequently, the rank does not increase along the~\eqref{eq:flowG}, which in turn allows us to successfully invoke the restricted injectivity of $\A$.}\hfill$\blacksquare$

 {In words,} Lemma~\ref{lem:flowNearby}  establishes that  {the}~\eqref{eq:flowG} never escapes the $\rho_0$-neighborhood of the set~$\M_b$ in~\eqref{eq:manifold}.
 {Lemma~\ref{lem:flowNearby}} will shortly enable us to prove  the convergence of  {the}~\eqref{eq:flowG}, after  recalling the {\L}ojasiewicz’s Theorem  {below}~\cite{lojasiewicz1982trajectoires,kurdyka2000proof}. 

\begin{thm}[\textsc{{\L}ojasiewicz’s Theorem}]\label{thm:lojaThm}
If $h:\R^n \rightarrow\R$ is an analytic function and the curve 
$
[0,\infty) \rightarrow \R^n$, $t\rightarrow z(t)$
% \label{eq:curve}
% \end{align}
is bounded and solves the gradient flow $\dot{z}(t) = - \nabla h(z)$, then this curve  converges to an FOSP of~$h$.
% as $t\rightarrow\infty$.  
\end{thm}

  {To apply Theorem~\ref{thm:lojaThm},} note  that $G(U)$ in~\eqref{eq:fgDefn} is an analytic function of $U$ on~$\R^{d\times p}$. 
Recall  {also} from~\eqref{eq:manifold} and~\eqref{eq:flowRemainsCloseAlways}  that  {the}~\eqref{eq:flowG} is bounded.  
We can now apply Theorem~\ref{thm:lojaThm}, which asserts that the~\eqref{eq:flowG}  converges to an FOSP  of $G$,  {denoted here by $\ol{U}\in \R^{d\times p}$}, such that $\dist(\ol{U},\M_b) \le \rho_0$. 
% see~\eqref{eq:flowRemainsCloseAlways}. 
In fact, by Lemma~\ref{lem:noSpFOSP}, this FOSP $\ol{U}$  {has zero training error, that is,}
\begin{align}
\A(\ol{U}\, \ol{U}^\top) =b.
\label{eq:limitIsFeas}
\end{align}
 {Moreover,} in view of~\eqref{eq:rankInvEq} and after using the fact that $\{U: \rank(U)\le \rank(U_0)\}$ is a closed set,  {we find that} the limit point $\ol{U}$  {of the~\eqref{eq:flowG}} also satisfies 
\begin{align}
    \rank(\ol{U}) \le \rank(U_0).
    \label{eq:rankInvarianceSOSP}
\end{align}
 {Informally speaking, we have thus far  established that the~\eqref{eq:flowG} has a limit point $\ol{U}$ with zero training error, see~\eqref{eq:limitIsFeas}, and this limit point is   low-rank, see~\eqref{eq:rankInvarianceSOSP}. We  next establish below that~$\ol{U}$ also has zero test error.}

 {Let us recall from Lemma~\ref{lem:flowNearby} that $\rank(U_0)\ge \effrank(X^\n)$ and that Assumption~\ref{assumption:iso} holds with $r\ge 2
\cdot \rank(U_0)$.} 
 {It follows immediately} that any matrix $X$ that satisfies  $\rank(X)\le \rank(U_0)$ and $\A(X) = b = \A(X^\n)$ must be infinitesimally close to~$X^\n$ in~\eqref{eq:measOp}. 
% $X^\n$ in~\eqref{eq:model} is the only matrix with rank at most $\rank(U_0)+r$ such that $\A(X^\n)=b$. 
In view of~\eqref{eq:limitIsFeas} and~\eqref{eq:rankInvarianceSOSP}, we therefore conclude that $\ol{U}\, \ol{U}^\top$ is infinitesimally close to $ X^\n$. 
% conclude that $\ol{U}\cdot \ol{U}^\top = X^\n$, 
In words,  {we conclude that the}~\eqref{eq:flowG} recovers the  {planted} matrix~$X^\n$,
 {up to an infinitesimal error.} 
  This completes the proof of Theorem \ref{thm:main}.

\section{Proof of Lemma \ref{lem:rankInvariance}}

For $U\in \R^{d\times p}$, recall the linear operator $\Der g(U)$ from~\eqref{eq:defnC}. The adjoint of this operator is $(\Der g(U))^*:\R^m\rightarrow\R^{d\times p}$, defined as
\begin{align}
    (\Der g(U))^* [\d] := \A^*(\d)\cdot U = \sum_{i=1}^m \d_i A_i U,
        \label{eq:adjO2}
\end{align}
where $\A^*$ is the adjoint of the linear operator $\A$ in~\eqref{eq:measOp} and $\d_i$ is the $i^{\text{th}}$ entry of the vector $\d\in \R^m$. 
For future use, recall also that
\begin{align}
    \nabla G(U_t) & = (\Der g(U_t))^* [g(U_t)] = \A^*(g(U_t))\cdot U_t,
    \qquad \text{(see \eqref{eq:fgDefn},\eqref{eq:adjO})}
    \label{eq:DerhGammaExplicitRecall}
\end{align}
where we used the shorthand $U_t = U(t)$. 
Let us also define
\begin{align}
X_t :=U_t U_t^\top \in \R^{d\times d},
\label{eq:xFlow}
\end{align}
and note that this new flow in $\R^{d\times d}$ satisfies
\begin{align}
   X_0 &= U_0 U_0^\top, \qquad \text{(see \eqref{eq:xFlow})} \nonumber\\
    \dot{X}_t & = \dot{U}_t U_t^\top + U_t \dot{U}_t^\top \qquad \text{(see \eqref{eq:xFlow})} \nonumber\\
    &= - \nabla G(U_t) \cdot U_t^\top  - U_t (\nabla G(U_t))^\top 
    \qquad \text{(see \eqref{eq:flowG})}
    \nonumber\\
    & = -\A^*(g(U_t)) X_t -  X_t \A^*(g(U_t)) .
    \qquad \text{(see (\ref{eq:DerhGammaExplicitRecall}),\eqref{eq:xFlow})}
    \label{eq:xFlowDer}
\end{align}
In the last identity above, we used the fact that $\{A_i\}_i$ are symmetric matrices in~\eqref{eq:measOp}. 
The next technical result establishes that the flow~\eqref{eq:xFlowDer} has an analytic SVD. 
\begin{lem}[\textsc{Analytic SVD}]\label{lem:analyticSVDx} 
% Under the assumptions made in Lemma~\ref{lem:rankInvariance}, 
The flow~\eqref{eq:xFlowDer} has the analytic SVD
\begin{align}
    X_t \overset{\text{SVD}}{=} V_t S_t V_t^\top,
    \qquad  t\ge 0,
    \label{eq:analyticSVDx}
\end{align}
where $V_t\in \R^{d\times d}$ is an orthonormal basis and the diagonal matrix $S_t\in \R^{d\times d}$ contains the singular values of $X_t$ in no particular order. Moreover, $V_t$ and $S_t$ are analytic functions of~$t$ on $[0,\infty)$. 
\end{lem}

\emph{Proof.} In view of~\eqref{eq:fgDefn},  $G(U)$ is an analytic function of $U$ in $\R^{d\times p}$. It then follows from Theorem~1.1 in~\cite{2008lectures} that the flow~\eqref{eq:flowG} is an analytic function of $t$ on $[0,\infty)$. Consequently, $X_t = U_t U_t^\top$ is an analytic function of $t$ on  $[0,\infty)$, see~\eqref{eq:xFlow}. It finally follows from Theorem~1 in~\cite{bunse1991numerical} that $X_t$ thus has an analytic SVD on $[0,\infty)$, as claimed. This completes the proof of Lemma~\ref{lem:analyticSVDx}.\hfill$\blacksquare$

% By comparing~(\ref{eq:xFlow}) and~(\ref{eq:analyticSVDx}), we record that 
% \begin{align}
%     U_t = V_t \sqrt{S_t} R_t,
%     \qquad  t \in [0,\tau],
%     \label{eq:uToV}
% \end{align}
% where $R_t\in \R^{d\times p}$ has orthonormal columns, i.e., $R_t^\top R_t = I_p$.\footnote{Indeed, let $U_t= V_t Q_t$ for a matrix $Q_t\in \R^{d\times p}$. It follows from~(\ref{eq:xFlow}) and~(\ref{eq:analyticSVDx}) that $Q_t Q_t^\top = S_t$, where we also  used the fact that $V_t$ is an orthonormal basis. It follows that $(\sqrt{S_t})^\dagger Q_t Q_t^\top (\sqrt{S_t})^\dagger = I_{S_t}$, where $I_{S_t}\in \R^{d\times d}$ is a diagonal matrix with nonzero entries, equal to one, corresponding to the nonzero entries of $S_t$. We observe that the left-hand side of last identity is the orthogonal projection matrix onto the canonical subspace corresponding to the nonzero diagonal entries of $S_t$. It follows that $(\sqrt{S_t})^\dagger Q_t=R_t$, where $R_t\in \R^{d\times p}$ has orthonormal columns and, consequently, $Q_t = \sqrt{S_t} R_t$ and, finally, $U_t = V_t \sqrt{S_t} R_t$, as claimed in~\eqref{eq:uToV}.     } 
By taking the derivative with respect to $t$ of both sides of~\eqref{eq:analyticSVDx}, we find that 
\begin{align}
    \dot{X}_t = \dot{V}_t S_t V_t^\top + V_t \dot{S}_t V_t^\top + V_t S_t \dot{V}_t^\top, 
    \qquad t\ge 0.
\end{align}
By multiplying both sides above by $V_t^\top$ and $V_t$ from left and right, we reach
\begin{align}
    V_t^\top \dot{X}_t V_t & = V_t^\top \dot{V}_t S_t + \dot{S}_t + S_t \dot{V}_t^\top V_t,
    \qquad t \ge 0,
    \label{eq:preSingEvolveX}
\end{align}
where we used on the right-hand side above the fact that $V_t$ is an orthonormal basis, i.e., $V_t^\top V_t = I_d$. Taking derivative of both sides of the last identity also yields that 
\begin{align}
    \dot{V}_t^\top V_t + V_t^\top \dot{V}_t = 0, 
    \qquad t \ge 0,
\end{align}
i.e., $V_t^\top \dot{V}_t$ is a skew-symmetric matrix. In particular, both $\dot{V}_t^\top V_t$ and $V_t^\top \dot{V}_t$ are hollow matrices, i.e., with zero diagonal entries. By taking the diagonal part of both sides of~\eqref{eq:preSingEvolveX}, we therefore arrive at 
\begin{align}
    \dot{s}_{t,i} & = v_{t,i}^\top \dot{X}_t v_{t,i},
    \qquad t \ge 0,
    \label{eq:singEvolvRawe}
\end{align}
where $s_{t,i}$ is the $i^{\text{th}}$ singular value of $X_t$ and $v_{t,i}\in \R^d$ is the corresponding singular vector.  By substituting above the  expression for $\dot{X}_t$ from~\eqref{eq:xFlowDer}, we find that 
\begin{align}
    \dot{s}_{t,i} 
    & =
    - 2 s_{t,i}\cdot v_{t,i}^\top \A^*(g(U_t)) v_{t,i}, \qquad t\ge 0, \qquad \text{(see (\ref{eq:xFlowDer}),(\ref{eq:singEvolvRawe}))}
    \label{eq:evolvSing}
\end{align}
where above we used the fact that $(s_{t,i},v_{t,i})$ is a pair of singular value and its corresponding singular vector for $X_t$.
In view of the evolution of singular values given by~\eqref{eq:evolvSing}, it is evident that
\begin{align}
    \rank(U_t) = \rank(X_t) \le \rank(X_0) = \rank(U_0), 
    \qquad t \ge 0,
    \qquad \text{(see \eqref{eq:xFlow})}
\end{align}
which completes the proof of Lemma~\ref{lem:rankInvariance}. The two identities above follow from~\eqref{eq:xFlow}. 
% We also remark that the choice of $\rho_0/2$ in Lemma~\ref{lem:rankInvariance} is cosmetic and can be replaced with any positive scalar strictly less than $\rho_0$. 

% \section{Proof of Lemma \ref{lem:analyticSVDx}}

% For $u\in \R^{d\times p}$ such that $\dist(u,\M_b) \le \rho$, recall the expression for $h_\g(u)$ as 
% \begin{align}
%     h_\g(u) & = L_\g(u,\lambda(u,u)) 
%     \qquad \text{(see \eqref{eq:hGamma})}
%     \nonumber\\
%     & = \frac{1}{2}\|u\|_\F^2 - \frac{1}{2} \langle A(uu^\top) - b, \lambda(u,u) \rangle + \frac{\g}{8}\|A(uu^\top) - b\|_2^2,
%     \qquad \text{(see (\ref{eq:al}))}
%     \nonumber\\
%     \lambda(u,u) & = K(u)^\dagger A(uu^\top) = K(u)^{-1} A(uu^\top).
%     \qquad \text{(see \eqref{eq:defnKernel})}
%     \label{eq:hGammaFullExp}
% \end{align}
% Above, the last identity above uses the fact that $\rank(K(u)) = m'=m$ by Lemma~\ref{lem:derLambda}, i.e., $K(u)$ is full-rank and thus invertible. 

% Since $m'=m$ by assumption, 
% In view of~\eqref{eq:fgDefn},  $G(U)$ is an analytic function of $U$ in $\R^{d\times p}$. It then follows from Theorem~1.1 in~\cite{2008lectures} that the flow~\eqref{eq:flowG} is an analytic function of $t$ on $[0,\infty)$. Consequently, $X_t = U_t U_t^\top$ is an analytic function of $t$ on  $[0,\infty)$, see~\eqref{eq:xFlow}. It finally follows from Theorem~1 in~\cite{bunse1991numerical} that $X_t$ thus has an analytic SVD on $[0,\infty)$, as claimed. This completes the proof of Lemma~\ref{lem:analyticSVDx}.  

\section{Proof of Lemma~\ref{lem:flowNearby}}\label{sec:proofThm}

 {In view of Definition~\ref{defn:effRank},} 
recall $X^\n$ in~\eqref{eq:model}  and fix $X^\n_l$  such that 
 \begin{align}
 \rank(X^\n_l) = \effrank(X^\n),
 \qquad 
 \|X^\n - X^\n_l\|_\F \le \epsilon \| X^\n\|_\F,
 \label{eq:Xlproofs}
\end{align}
for an infinitesimal $\epsilon$. 
Also recall from Assumption~\ref{assumption:key}\ref{assumption:infoLevel} that $ \rank(X^\n) \le p$ and let us~fix $U^\n\in \R^{d\times p}$ such that
\begin{align}
U^\n (U^\n)^\top = X^\n.
\label{eq:UshToXl}
\end{align}
 {Using the shorthand $U_t=U(t)$,} we then note that 
\begin{align}
 \|g(U_0)\|_2 & \ge \|g(U_t)\|_2   \qquad \text{(see \eqref{eq:fgDefn},\eqref{eq:flowG})} \nonumber\\
& = \frac{1}{2}\|\A(U_t U_t^\top) - b\|_2 
  \qquad \text{(see \eqref{eq:fgDefn})}  \nonumber\\
  & =
  \frac{1}{2} \| \A( U_t U_t^\top - X^\n )\|_2 
  \qquad \text{(see \eqref{eq:measOp})}\nonumber\\
  & \ge   \frac{1}{2} \| \A( U_t U_t^\top - X^\n_l )  \|_2 - \frac{1}{2}\|\A( X^\n - X^\n_l )\|_2 \qquad \text{(triangle inequality)} \nonumber\\
  & \ge \frac{1}{2} \| \A( U_t U_t^\top - X^\n_l )  \|_2 - \epsilon, 
  \qquad \text{(see \eqref{eq:Xlproofs})}
%   \frac{\epsilon\|\A\|\|X^\n\|_\F}{2},
  \label{eq:beforeRIPFlow}
\end{align}
for every $t\ge 0$, where $\epsilon$ is infinitesimal. As seen above, throughout this proof, we will keep the notation light by absorbing all  {bounded} factors into  {the infinitesimal} $\epsilon$.  The argument of $\A(\cdot)$ in the last line above is  low-rank,  {that is}
\begin{align}
    \rank(U_t U_t^\top - X_l^\n) &  \le \rank(U_t U_t^\top) + \rank(X^\n_l) \nonumber\\
    & = \rank(U_t) + \effrank(X^\n)  
    \qquad \text{(see \eqref{eq:Xlproofs})}
    \nonumber\\
    & \le  \rank(U_0 ) + \effrank(X^\n),
    \qquad \text{(see Lemma~\ref{lem:rankInvariance})}
    \nonumber\\
    & \le 2\cdot \rank(U_0),
\end{align}
for every $t\ge 0$. The last line above uses the assumption that $\effrank(X^\n) \le \rank(U_0)$. 
 {On the other hand},  {also} by assumption, the linear operator $\A$ satisfies the {$(2\cdot \rank(U_0))$}-RIP, which allows us to lower bound the last line of~\eqref{eq:beforeRIPFlow} as 
\begin{align}
    \|g(U_0)\|_2 & \ge \frac{1}{2} \| \A( U_t U_t^\top - X^\n_l )\|_2 - \epsilon \qquad \text{(see \eqref{eq:beforeRIPFlow})} \nonumber\\
    & \ge \frac{\a}{2} \|U_t U_t^\top - X^\n_l\|_\F - \epsilon
    \qquad \text{(see Assumption~\ref{assumption:iso})}
    \nonumber\\
    & \ge \frac{\a}{2} \|U_t U_t^\top - X^\n\|_\F - \frac{\a}{2}\| X^\n - X^\n_l\|_\F - \epsilon\qquad \text{(triangle inequality)} \nonumber\\
    & \ge \frac{\a}{2} \|U_t U_t^\top - X^\n\|_\F - \epsilon
    \qquad \text{(see \eqref{eq:Xlproofs})}
    \nonumber\\
    & = \frac{\a}{2} \|U_t U_t^\top - U^\n (U^\n)^\top\|_\F - \epsilon,
    \qquad \text{(see \eqref{eq:UshToXl})}
    \label{eq:postRIPFlow}
\end{align}
for every $t\ge 0$. Above,  $\a>0$ is a constant,  {which is sometimes referred to as the isometry constant of $\A$,} and $\epsilon$ is infinitesimal. 
To  lower bound the last term above, we rely on the following technical lemma. 

\begin{lem}[\textsc{Generalized orthogonal Procrustes problem}]\label{lem:procrustes}
    For matrices $U,V\in \R^{d\times p}$, 
    % let $p'=\max(\rank(U),\rank(v))$. Then 
    it holds that 
    \begin{align}
        \| UU^\top - V V^\top \|_\F & 
        % \ge \min\l(\s_{\min}(U), \s_{\min}(\M_V)\r)  \cdot \dist(U, V \O_p) \nonumber\\
        \ge \max\l(\s_{\min}(U), \s_{\min}(\M_V)\r)  \cdot \dist(U, \M_V),
        % (\s_{p'}(u) + \s_{p'}(v) ) \cdot \dist(u, v \O_{p'}) \ge \s_{\min}(\M_v)  \cdot \dist(u, \M_v) ,
    \end{align}
    % where 
    % \begin{align}
    %     \dist(u, v\O_{p'}) := \min_{R\in \O_{p'}}\,\,\| u - vR\|_\F,
    % \end{align}
    % is the orthogonal Procrustes distance, and
    % $\O_{p'}$ is the orthogonal group, i.e., $\O_{p'} = \{ R: R^\top R = I_{p'}\}$. Moreover, 
    % where $V\O_p := \{ VR: R^\top R = I_p \} \subset \R^{d\times p}$ and 
    where 
    $$ 
    \M_V := \{V' : \A(V'V'^\top) = \A(VV^\top),\, \|V'\| \le \xi \} \subset \R^{d\times p},
    $$
        \begin{align}
        \s_{\min}(\M_V) := \min_{V'\in \M_V} \s_{\min}(V').
    \end{align}
    Above, $\s_{\min}(U)$ and $\s_{\min}(V')$ are the smallest nonzero singular values of $U$ and $V'$, respectively. 
\end{lem}

In order to apply Lemma~\ref{lem:procrustes}  {to the last line of~\eqref{eq:postRIPFlow}}, suppose that  {the}~\eqref{eq:flowG} is initialized at $U_0\in \R^{d\times p}$ such that $\dist(U_0,\M_b) <\rho_0$ and let $\tau\in (0,\infty) $ (if it exists) denote the smallest number such that $\dist(U_t,\M_b) = \rho_0$. In particular, note that  {the}~\eqref{eq:flowG} is initialized and remains in the $\rho_0$-neighborhood of the set $\M_b$ for every $t\le \tau$. That is, $\dist(U_t,\M_b) \le \rho_0 $ for every $t\le \tau$.
% Recall also that Assumption~\ref{assumption:key} holds with $m'=m$. 
We now apply Lemma~\ref{lem:procrustes} to the last line of~\eqref{eq:postRIPFlow} to reach
\begin{align}
    \|g(U_0)\|_2 & \ge \frac{\a}{2} \|U_t U_t^\top - U^\n (U^\n)^\top\|_2 - \epsilon
    \qquad \text{(see \eqref{eq:postRIPFlow})}
    \nonumber\\
    & \ge \frac{\a}{2}  \s_{\min}(\M_b) \dist(U_t, \M_b) - \epsilon,\nonumber\\
    & \qquad\qquad \qquad \text{(see \eqref{eq:measOp},\eqref{eq:manifold},\eqref{eq:UshToXl} and Lemma~\ref{lem:procrustes})}
    % & \ge \frac{\a}{2} \cdot \s_{\min}(\M_{b,\rho_0}) \dist(U_t, \M_b)  - \epsilon,
    \label{eq:beforeHowSmall}
\end{align}
for every $t\le \tau$.
% where  {we defined}
% \begin{align}
%     \s_{\min}(\M_{b,\rho_0}) := \min\l\{ \s_{\min}(U) : \dist(U,\M_b) \le \rho_0 \r\} >0.
%     \label{eq:sMinNeigh}
% \end{align}
% Above, the minimum is achieved and is positive because $\M_b$ in~\eqref{eq:manifold} is compact and $\s_{\min}(\cdot)$ is continuous. 
 {To obtain a more informative result,}
we next  upper bound $\|g(U_0)\|_2$ in~\eqref{eq:beforeHowSmall} as follows.  {First,} let $U_{0,\M_b}\in \M_b$ denote the projection of $U_0$ on $\M_b$ in~\eqref{eq:manifold},  {that is,}
\begin{align}
    \dist(U_0,\M_b) & = \|U_0 - U_{0,\M_b}\|_2 \le \|U_0 - U'\|_2,\qquad \text{if } U'\in \M_b. 
    \label{eq:projInit}
\end{align}
Above, by compactness of $\M_b$ in~\eqref{eq:manifold}, the projection $U_{0,\M_b}$ exists but might not be unique. Using~\eqref{eq:projInit}, we then upper bound $\|g(U_0)\|_2$ as 
\begin{align}
    \|g(U_0)\|_2 & = \frac{1}{2} \| \A(U_0U_0^\top) - b\|_2 
    \qquad \text{(see \eqref{eq:fgDefn})}
    \nonumber\\
    & = \frac{1}{2} \| \A(U_0U_0^\top- U_{0,\M_b}U_{0,\M_b}^\top ) \|_2
    \qquad \text{(see \eqref{eq:projInit}, {\eqref{eq:manifold}})} 
    \nonumber\\
    & \le \frac{1}{2} \|\A\|\cdot  \|  U_0U_0^\top - U_{0,\M_b}U_{0,\M_b}^\top\|_\F 
     \nonumber\\
    & \le \frac{1}{2}\|\A\|\cdot  (\|U_0\|+\|U_{0,\M_b}\|) \cdot \|U_0 - U_{0,\M_b}\|_\F \nonumber\\
    & \le \|\A\|\xi \cdot  \|U_0 - U_{0,\M_b}\|_\F \nonumber\\
    & = \|\A\|\xi \cdot  \dist(U_0,\M_b),
    \qquad \text{(see \eqref{eq:projInit})}
    \label{eq:lowBndGU0}
\end{align}
where the second-to-last line above assumes that $\xi$ satisfies 
\begin{align}
\xi \ge \|U_0\|.
\label{eq:xiLargeEnough}
\end{align}
The second-to-last line in~\eqref{eq:lowBndGU0} also uses the fact that  $U_{0,\M_b}\in \M_b$  satisfies $\|U_{0,\M_b}\|\le \xi$, see~\eqref{eq:manifold}. By combining the  lower and upper bounds for $\|g(U_0)\|_2$ in~\eqref{eq:beforeHowSmall} and~\eqref{eq:lowBndGU0}, we  {finally} arrive at
\begin{align}
    \dist(U_t, \M_b) &
    \le \frac{2}{\alpha \sigma_{\min}(\M_{b}) }\|g(U_0)\|_2 + \epsilon
    \qquad \text{(see \eqref{eq:beforeHowSmall})}
    \nonumber\\
    &\le \frac{2\xi \|\A\| }{\a \cdot \s_{\min}(\M_{b}) } \dist(U_0,\M_b) + \epsilon,
    \qquad \text{(see \eqref{eq:lowBndGU0})}
    \label{eq:beforeHowSmallNew}
\end{align}
for every $t\le \tau$, where $\epsilon$ is infinitesimal. 
By setting 
\begin{align}
     \dist(U_0,\M_b) <\rho_0 \min\l( 1,\frac{\a  \cdot \s_{\min}(\M_{b}) }{4 \xi \|\A\|} \r) -\epsilon,
\label{eq:howSmallInit}
\end{align}  
for an infinitesimal $\epsilon$, 
it follows from~\eqref{eq:beforeHowSmallNew} that  $\dist(U_t,\M_b)< \rho_0 $ for every $t\le \tau$. Recalling the definition of $\tau$ earlier,  {to avoid the contradiction}, we conclude that $\dist(U_t,\M_b)<\rho_0$ holds for every $t\ge 0$.  
This completes the proof of Lemma~\ref{lem:flowNearby}.  {Alternatively, in view of~\eqref{eq:beforeHowSmallNew}, we can set
\begin{align}
     \|\A(U_0U_0^\top) - b\|_2 = 2\|g(U_0)\|_2 < \rho_0 \a  \cdot \s_{\min}(\M_{b})  -\epsilon,
\label{eq:howSmallInitV2}
\end{align}  
instead of~\eqref{eq:howSmallInit} and then arrive at the same conclusion that $\dist(U_t,\M_b)<\rho_0$ holds for every $t\ge 0$. Likewise, since $\|\A(U_0U_0^\top) - b\|_2 \le \|\A\|\cdot \|U_0U_0^\top-X^\n\|$ by~\eqref{eq:measOp}, we can  set
\begin{align}
    \|U_0U_0^\top - X^\n\|_\F \le \frac{\rho_0\alpha}{\|\A\|}\cdot \s_{\min}(\M_{b}) - \epsilon,
\end{align}
and again arrive at the same conclusion that~$\dist(U_t,\M_b)<\rho_0$ for every~$t\ge 0$.
}

\section{Proof of Lemma \ref{lem:procrustes}}

To prove Lemma~\ref{lem:procrustes}, let us recall the following standard notion~\cite{higham1995matrix}. 
\begin{lem}[\textsc{Orthogonal Procrustes problem}] For matrices $U,V\in \R^{d\times p}$, the orthogonal Procrustes problem is solved as 
\begin{align}
        \dist(U, V\O_{p})= \dist(U\O_p, V) := \min_{R\in \O_{p}}\,\,\| U - VR\|_\F = \sqrt{\|U\|_\F^2 + \|V\|_\F^2 - 2 \|U^\top V\|_*},
        \label{eq:defnProcturusPr}
    \end{align}
where $\O_p$ is the orthogonal group, i.e., $V\O_p=\{R\in \R^{p\times p}: R^\top R = I_p\} \subset \R^{d\times p}$, and $\|\cdot\|_*$ stands for the nuclear norm.
\end{lem}
We also recall another standard result below~\cite[Lemma 3]{li2019non}, which is proved in Appendix~\ref{sec:genProcZZ} for completeness.
\begin{lem}[\textsc{Orthogonal Procrustes problem}]\label{lem:procrustes0} For matrices $U,V\in \R^{d\times p}$, it holds that 
 \begin{align}
        \| UU^\top - V V^\top \|_\F \ge \max(\s_{\min}(U),\s_{\min}(V)) \cdot \dist(U, V\O_{p}),
\end{align}
where $\s_{\min}(U)$ and $\s_{\min}(V)$ are the smallest nonzero singular values of $U$ and $V$, respectively.  
\end{lem}
On the other hand, because 
\begin{align}
V \O_p \subset \M_V := \{V': \A(V'V'^\top) = \A(VV^\top),\, \|V'\|\le  \xi\}, 
\label{eq:MVdefn}
\end{align}
it holds that 
\begin{align}
    \dist(U,V\O_p) \ge \dist(U,\M_V). 
    \label{eq:orderDists}
\end{align}
In combination with Lemma~\ref{lem:procrustes0},~\eqref{eq:orderDists} implies that 
\begin{align}
    \|UU^\top - VV^\top\|_\F & \ge \max(\s_{\min}(U),\s_{\min}(V)) \cdot \dist(U,V\O_p)
    \qquad \text{(see Lemma~\ref{lem:procrustes0})}
    \nonumber\\
    % & \ge \min(\s_{\min}(U),\s_{\min}(V)) \cdot  \dist(U,\M_V) 
    % \qquad \text{(see \eqref{eq:orderDists})} \nonumber\\
    & \ge \max\l( \s_{\min}(U), \min_{V'\in \M_V}\s_{\min}(V') \r) \cdot \dist(U,V \O_p) 
    \qquad (V\in \M_V)
    \nonumber\\
    & =: \max\l(\s_{\min}(U),  \s_{\min}(\M_V) \r) \cdot \dist(U,V\O_p) \nonumber\\
    & \ge \max\l(\s_{\min}(U),  \s_{\min}(\M_V) \r) \cdot \dist(U,\M_V),
    \qquad \text{(see \eqref{eq:orderDists})}
\end{align}
where the minimum  over the set $\M_V$ is achieved by the compactness of $\M_V$ in~\eqref{eq:MVdefn} and the continuity of $\s_{\min}(\cdot)$. 
This completes the proof of Lemma~\ref{lem:procrustes}.

\section{Proof of Lemma \ref{lem:procrustes0}}\label{sec:genProcZZ} For the convenience of the reader, the proof is repeated from~\cite[Lemma 3]{li2019non} after minor modifications.
Without loss of generality, we assume throughout this proof that $U$ and $V$ have the singular value decomposition of the form
\begin{align}
    U = \ol{U}\cdot  \Sigma_U, \qquad V = \ol{V}\cdot  \Sigma_V,
    \label{eq:wlog}
\end{align}
where $\ol{U},\ol{V}\in \R^{d\times d}$ are orthonormal bases, and the diagonal matrices $\Sigma_U,\Sigma_V\in \R^{d\times d}$ collect the singular values of $U$ and $V$, respectively. 
Note that 
\begin{align}
    \|UU^\top - VV^\top\|_\F^2 & = \|UU^\top \|_\F^2 + \|VV^\top \|_\F^2 - 2 \langle UU^\top , VV^\top \rangle \nonumber\\
    & = \trace(U U^\top U U^\top ) + \trace(V V^\top V V^\top ) - 2 \|U^\top V\|_\F^2 \nonumber\\
    & = \trace(\Sigma_U^4) + \trace(\Sigma_V^4) - 2 \| \Sigma_U \ol{U}^\top \ol{V} \Sigma_V \|_\F^2 \nonumber\\
    & = \sum_{i=1}^d \sigma_{U,i}^4 + \sum_{j=1}^d \s_{V,j}^4 - 2 \sum_{i,j=1}^d \s_{U,i}^2 \s_{V,j}^2 (\ol{u}_i^\top \ol{v}_j)^2, 
    \label{eq:traceBack1}
\end{align}    
where $\s_{U,i}$ and $\s_{V,j}$ are the $i^{\text{th}}$ and $j^{\text{th}}$ diagonal entries of $\Sigma_U$ and $\Sigma_V$, respectively. Also, $\ol{u}_i$ and $\ol{v}_j$ above are the $i^{\text{th}}$ and $j^{\text{th}}$ columns of $\ol{U}$ and $\ol{V}$, respectively. Since $\ol{U}$ and $\ol{V}$ both have orthonormal columns, we rewrite the last line above as
\begin{align}
    & \sum_{i=1}^d \sigma_{U,i}^4 + \sum_{j=1}^d \s_{V,j}^4 - 2 \sum_{i,j=1}^d \s_{U,i}^2 \s_{V,j}^2 (\ol{u}_i^\top \ol{v}_j)^2 \nonumber\\
    & = \sum_{i=1}^d \sigma_{U,i}^4 \|\ol{u}_i\|_2^2 + \sum_{j=1}^d \s_{V,j}^4 \|\ol{v}_i\|_2^2 - 2 \sum_{i,j=1}^d \s_{U,i}^2 \s_{V,j}^2 (\ol{u}_i^\top \ol{v}_j)^2 \nonumber\\
    & = \sum_{i,j=1}^d \s_{U,i}^4 (\ol{u}_i^\top \ol{v}_j)^2+\sum_{i,j=1}^d \s_{V,j}^4 (\ol{u}_i^\top \ol{v}_j)^2  - 2 \sum_{i,j=1}^d \s_{U,i}^2 \s_{V,j}^2 (\ol{u}_i^\top \ol{v}_j)^2.
    \label{eq:traceBack2}
\end{align}
We can rewrite the last line above more compactly as    
\begin{align}    
    & \sum_{i,j=1}^d \s_{U,i}^4 (\ol{u}_i^\top \ol{v}_j)^2+\sum_{i,j=1}^d \s_{V,j}^4 (\ol{u}_i^\top \ol{v}_j)^2  - 2 \sum_{i,j=1}^d \s_{U,i}^2 \s_{V,j}^2 (\ol{u}_i^\top \ol{v}_j)^2 \nonumber\\
    & =  \sum_{i,j=1}^d \l( \s_{U,i}^4 + \s_{V,j}^4   - 2  \s_{U,i}^2 \s_{V,j}^2 \r) (\ol{u}_i^\top \ol{v}_j)^2 \nonumber\\
    & = \sum_{i,j=1}^d (\s_{U,i}^2-\s_{V,j}^2)^2 (\ol{u}_i^\top \ol{v}_j)^2 \nonumber\\
    & = \sum_{i,j=1}^d (\s_{U,i}^2-\s_{V,j}^2)^2 (\ol{u}_i^\top \ol{v}_j)^2 \nonumber\\
    & \ge \sum_{i,j=1}^d (\s_{U,i}+\s_{V,j})^2 (\s_{U,i}- \s_{V,j})^2 (\ol{u}_i^\top \ol{v}_j)^2 \nonumber\\ 
    & \ge \l(\max(\s_{\min}(U),\s_{\min}(V))\r)^2 \sum_{i,j=1}^d (\s_{U,i}-\s_{V,j})^2 (\ol{u}_i^\top \ol{v}_j)^2,
    \label{eq:preRelate}
\end{align}
where $\s_{\min}(U)$ and $\s_{\min}(V)$ are the smallest nonzero singular value of $U$ and $V$, respectively. By tracing back our steps in~(\ref{eq:traceBack1})-(\ref{eq:preRelate}), we find that 
\begin{align}
    & \|UU^\top - VV^\top \|_\F^2  \nonumber\\
    & \ge (\max(\s_{\min}(U),\s_{\min}(V)))^2 \l(\sum_{i,j=1}^d (\s_{U,i}-\s_{V,j})^2 (\ol{u}_i^\top \ol{v}_j)^2\r)  \qquad \text{(see (\ref{eq:traceBack1})-(\ref{eq:preRelate}))}
    \nonumber\\
    & = (\max(\s_{\min}(U),\s_{\min}(V)))^2  \l( \|U\|_\F^2 + \|V\|_\F^2 - 2 \l\langle \Sigma_U \ol{U}^\top \ol{V} \Sigma_V, \ol{U}^\top \ol{V} \r\rangle \r).
    \label{eq:preRelate2}
\end{align}
To relate the last line above to $\dist(U,V\O_p)$, we next relate the last inner product above to the nuclear norm in~\eqref{eq:defnProcturusPr}, i.e., $\|U^\top V\|_*$. To that end, note that 
\begin{align}
    \|U^\top V\|_* & = \max_{\|Q\|\le 1} \,\, \langle U^\top V, Q\rangle \nonumber\\
    & \ge \langle U^\top V, \ol{U}^\top \ol{V} \rangle
    \qquad (\|\ol{U}^\top \ol{V}\| \le \|\ol{U}\|\cdot \|\ol{V}\| = 1 \text{ by \eqref{eq:defnProcturusPr}}) \nonumber\\
    & = \langle \Sigma_U \ol{U}^\top \ol{V} \Sigma_V,  \ol{U}^\top \ol{V}. \rangle,
    \qquad \text{(see \eqref{eq:wlog})}
    \label{eq:relNucFrob}
\end{align}
By substituting the above bound back into~\eqref{eq:preRelate}, we find that 
\begin{align}
    & \|UU^\top - VV^\top\|_\F^2 
    \nonumber\\
    & \ge 
    (\max(\s_{\min}(U),\s_{\min}(V)))^2 
    \l( \|U\|_\F^2 + \|V\|_\F^2 - 2 \l\langle \Sigma_U \ol{U}^\top \ol{V} \Sigma_V, \ol{U}^\top \ol{V} \r\rangle \r)
    \quad \text{(see \eqref{eq:preRelate2})}
    \nonumber\\
    & \ge (\max(\s_{\min}(U),\s_{\min}(V)))^2  \l(\|U\|_\F^2+ \|V\|_\F^2 - 2 \|U^\top V\|_* \r)  \qquad \text{(see \eqref{eq:relNucFrob})} \nonumber\\
    & = (\max(\s_{\min}(U),\s_{\min}(V)))^2  \cdot (\dist(U, V\O_p))^2,
    \qquad \text{(see \eqref{eq:defnProcturusPr})}
\end{align}
which completes the proof of Lemma~\ref{lem:procrustes0}.

\end{document}